\documentclass[%
 reprint,
superscriptaddress,
 amsmath,amssymb,
 pre,
]{revtex4-2}
\usepackage{centernot}
\usepackage{graphicx}
\usepackage{dcolumn}
\usepackage{bm}
\usepackage{times}
\usepackage{algpseudocodex}
\usepackage{algorithm}

\begin{document}


\title{Polarisation in increasingly connected societies}

\author{Tuan Minh Pham}
\email[Contact author: ]{m.t.pham@uva.nl} 
\affiliation{Dutch Institute for Emergent Phenomena, 1090 GE, Amsterdam, The Netherlands}
\affiliation{Institute for Advanced Study,
Oude Turfmarkt 147, 1012 GC Amsterdam, The Netherlands}
\affiliation{Institute of Physics, University of Amsterdam, Science Park 904, Amsterdam, The Netherlands}
\author{Sidney Redner}
\affiliation{Santa Fe Institute, 1399 Hyde Park Road, Santa Fe, NM 87501, USA}

\author{Lourens Waldorp}
\affiliation{Dutch Institute for Emergent Phenomena, 1090 GE, Amsterdam, The Netherlands}
\affiliation{University of Amsterdam, Nieuwe Achtergracht 129-B, Amsterdam 1018 NP, The Netherlands}

\author{Jay Armas}
\affiliation{Dutch Institute for Emergent Phenomena, 1090 GE, Amsterdam, The Netherlands}
\affiliation{Institute for Advanced Study,
Oude Turfmarkt 147, 1012 GC Amsterdam, The Netherlands}
\affiliation{Institute of Physics, University of Amsterdam, Science Park 904, Amsterdam, The Netherlands}
\affiliation{The Niels Bohr Institute, University of Copenhagen, Blegdamsvej 17, DK-2100, Denmark}

\author{Han L. J. van der Maas}
\affiliation{Dutch Institute for Emergent Phenomena, 1090 GE, Amsterdam,
The Netherlands}
\affiliation{University of Amsterdam, Nieuwe Achtergracht 129-B, Amsterdam 1018 NP, The Netherlands}

\begin{abstract}
 Explanations of polarization often rely on one of the three mechanisms: homophily, bounded confidence, and community-based interactions. Models based on these mechanisms consider the lack of interactions as the main cause of polarization. Given the increasing connectivity in modern society, this explanation of polarization may be insufficient. We aim to show that in involvement-based models, society becomes more polarized as its connectedness increases. To this end, we propose a minimal voter-type model (called I-voter) that incorporates involvement as a key mechanism  in opinion formation and study its dependence on the network connectivity. We describe the  steady-state behaviour of the model  analytically,  at the mean-field  and the moment-hierarchy levels and  stress the generality of our findings  by considering various  extensions and different network topologies.
\end{abstract}

\maketitle

\section{Introduction}
Polarization is a complex social phenomenon with likely multiple causes. While its study traces back to the 19th century \cite{Durkheim}, the field has grown rapidly over the past 30 years, driven by the availability of large-scale (social media) data and accelerated theoretical developments \cite{Dimant}.
While not the only mechanism (see \cite{Peralta2022} for a review of various explanations \footnote{Other models of polarization aim to capture phenomena like opinion distancing \cite{Jager2005}, persuasive arguments \cite{Mas2013}, social feedback between internal and expressed opinions \cite{Banisch2019, Ben2024}, biased assimilation \cite{Dandekar5791}, algorithmic mediation \cite{Peralta2021}, multidimensional (i.e. context-dependent) polarisation \cite{BaumannPRX,Peralta}, and conflict \cite{Tornberg}.}), many models emphasize the pivotal role of limited interactions, isolation, and, more generally, low connectivity in driving polarization .

For instance, in classical consensus models, such as DeGroot’s model \cite{DeGroot}, Abelson’s model \cite{abelson1964}, and the voter model \cite{Holley}, polarization arises when subgroups become disconnected, preventing the formation of a unified consensus \cite{Castellano2009}. Also the model of Axelrod \cite{RAxelrod1997, Flache2011,lanchier2012} predicts that, due to homophily,  small societies fragment into cultural groups at a critical number of alternative traits per feature. In this model, individuals are more likely to interact with “similar” neighbours than with dissimilar ones, and they become more similar after every interaction. 
Other models show how fragmentation results from the
presence of  “boundedly confident” agents \cite{Deffuant, Hegselmann2002, BERNARDO2024} who only interact with those not further away than a given distance in the opinion space. In threshold type models, agents only adopt a view once the fraction of neighbors supporting the same view exceeds their own threshold drawn from a predefined distribution of adoption thresholds \cite{Granovetter1978, Watts2002, Centola2007}. Polarisation can also emerge from rearrangements of social ties in co-evolving networks to form sparsely-connected \cite{Holme, Vazquez2008, Durrett2012, Baumann} or even antagonistic clusters of individuals \cite{Altafini2013, FLACHE20112, Pham2020, Gorski2023}. 

In these models, polarization typically arises from a lack of connectivity among agents holding opposing or distant opinions. However, in many respects, connectivity has increased in modern societies. Urbanization, globalization, international mobility, the rise of social media platforms, and cross-cultural marriages \cite{Smith2024, Kazmina} have all contributed to humans being far more interconnected today than a century ago.

In this paper, we explore a new explanation for polarization in increasingly interconnected societies, emphasizing the pivotal role of involvement in the process of opinion formation. The role of involvement (defined as sustained attention) has been extensively studied in psychology \cite{Blair}. It has been used to explain why attitudes and opinions sometimes behave like dimensions and sometimes act as categories {\cite{ lataneAttitudesCatastrophesDimensions1994a}. In network models of attitudes {\cite{Dalege2018} involvement plays a similar role. Hoffstadt et al. \cite{Hoffstadt} shows in several datasets that involvement increases poralization. Central aspects of involvement in opinion or attitude  formation include: low-involvement attitudes are more situationally influenced and less stable \cite{Petty1986}, when people feel highly involved, their attitudes are less sensitive to persuasion \cite{petty2014attitude}, involvement weakens over time when not reinforced \cite{Richins}  but can  sharply increase due to interactions. 

Building on these psychological models and empirical findings, we propose incorporating involvement into models of polarization. An earlier model developed along these lines—the HIOM model \cite{HIOM}—yielded several unexpected results, suggesting that polarization can emerge in highly connected networks even when all agents are, on average, receive the same information from their social  milieux. The findings from the HIOM model, however, were based solely on simulations and possibly confounded by other factors. In this paper, we aim to develop a minimal model to study the principal effects of involvement on polarization analytically.

Our setup is based on the well-studied constrained 3-state voter model \cite{Vazquez2004, Mobilia2011}. In \cite{Vazquez2004}, agents can be in one of three states: leftist, centrist, and rightist; and can only switch to neighboring states (e.g., leftist to centrist or rightist to centrist) but not directly between the extremes (e.g., leftist to rightist)\footnote{This latter assumption can be considered as the restriction of a more general scenario, where, at high network connectivity,  leftists and rightists can frequently interact with each other, but  rarely update their respective opinions.}.
The outcome of the constrained 3-state voter dynamics  can be either consensus or polarization with a frozen mixture of leftists and rightists. Here we  generalise this model by  assuming  extreme agents (either leftists or rightists) (a) when being in isolation, can turn into neutral ones with a nonvanishing rate; (b) when engaging in discussion  with neutral agents, are less susceptible  to the argument of the latter and are more likely to persuade the latter to become extreme. The resulting model is what we refer to as the I-voter model. Note that, the I-voter does not  employ any  mechanisms hindering consensus like other generalisations of the voter model (VM), such as individual stubbornness, partisanship, non-linear update, zealots (e.g. quenched disorder), or individual and social heterogeneity   \cite{Arkadiusz, Redner2019, Mobilia2007, Meyer2024, Khalil, DeMarzo, Starnini2012,Castellano_qvoter,Nyczka2012, Nowak2022}.  The relation to HIOM and other models is further discussed in the Appendix  \ref{related_model}.}

In the present paper, we formulate an analytical framework for a general network topology that allows for a mean-field treatment of the model's behavior and then demonstrate the validity of our approach on different network topologies. We will show that on sparse networks, the I-voter model yields increased polarisation with a growing number of interaction partners. We first describe and present results regarding the baseline model of three states and then provide a generalisation to the case of arbitrarily (odd) number of states.

\section{The model}
In the model, $N$ agents, each residing at a node in a social network, hold an opinion $x_i\in\{-1,0,1\}$ that stands for  “leftist”, “centrist” and “rightist”, respectively. When a leftist (rightist) and a centrist are in contact, the latter becomes left (right) with probability (per unit time) $p$, while an extremist turns centrist with probability $1-p$. 
The probability $p$ represents combined effects of interactions leading to persuasion, while $1-p$ represents reading or hearing other opinions that will make  someone's opinion more neutral.
As centrists are expected to be more easily influenced, their transition rate is necessarily larger than that of extremists, i.e.  $p >1-p$. We thus only consider the case of $p>0.5$. 
Furthermore, involvement (essentially attention) is a limited resource, meaning extremists may lose interest in the discussed issue and gradually become centrists, or there could be individual-specific effects, such as memory or disturbances (forgetting), or interference with other memories \cite{Murre,Edwards}. Thus the extremist, either left or right, can decay towards the center with probability (per unit time) $\epsilon$.  
While real-world agents are always subjected to stimulation, especially on topics that tend to polarize, 
allowing involvement to decay results in  a non-trivial opinion formation process that is not purely driven by  transitions  towards more extreme opinions. Figure 1 illustrates the dynamics of our I-voter model.

\begin{figure}
\centering
\includegraphics[width=.43\linewidth]{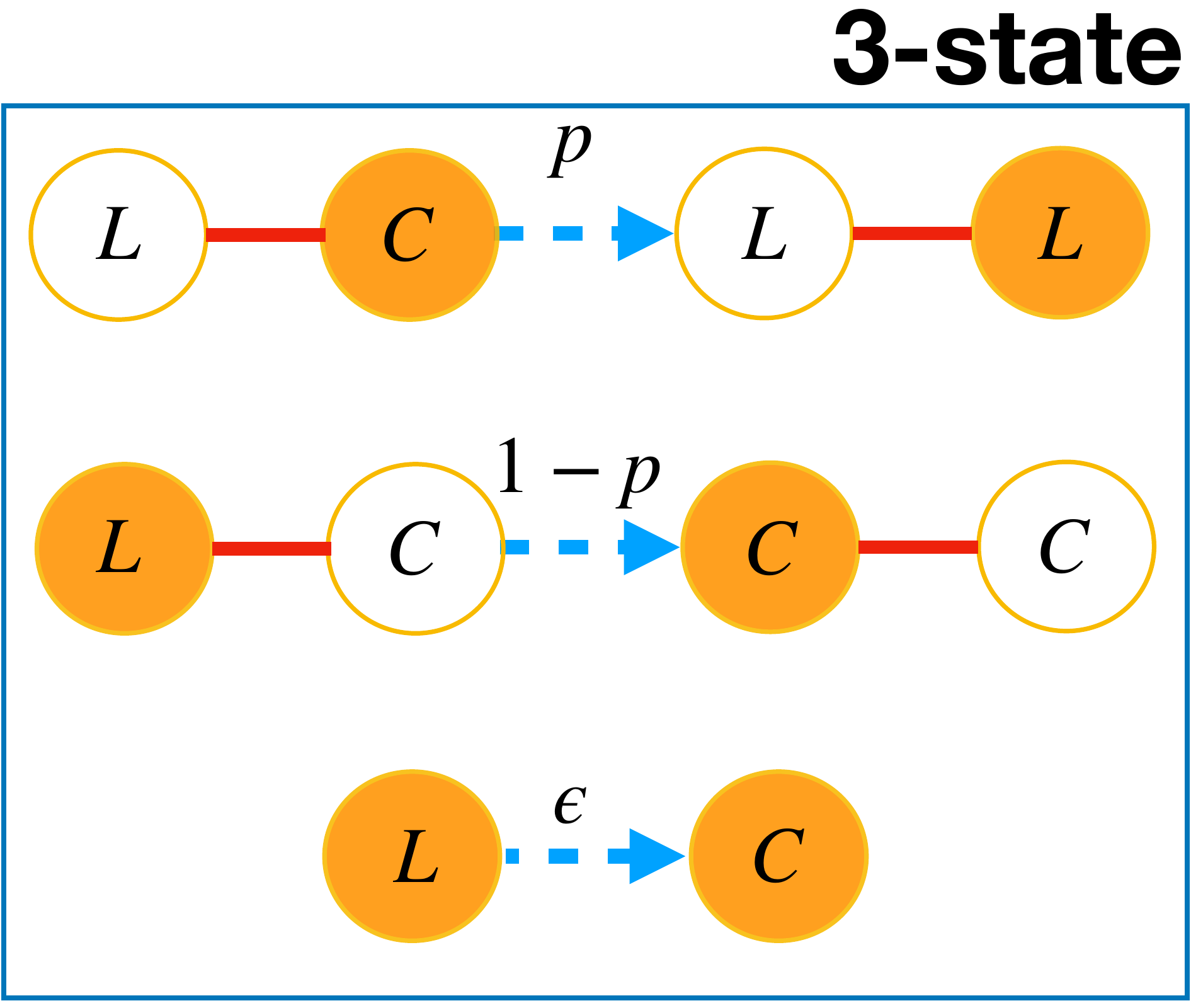} 
\includegraphics[width=.43\linewidth]{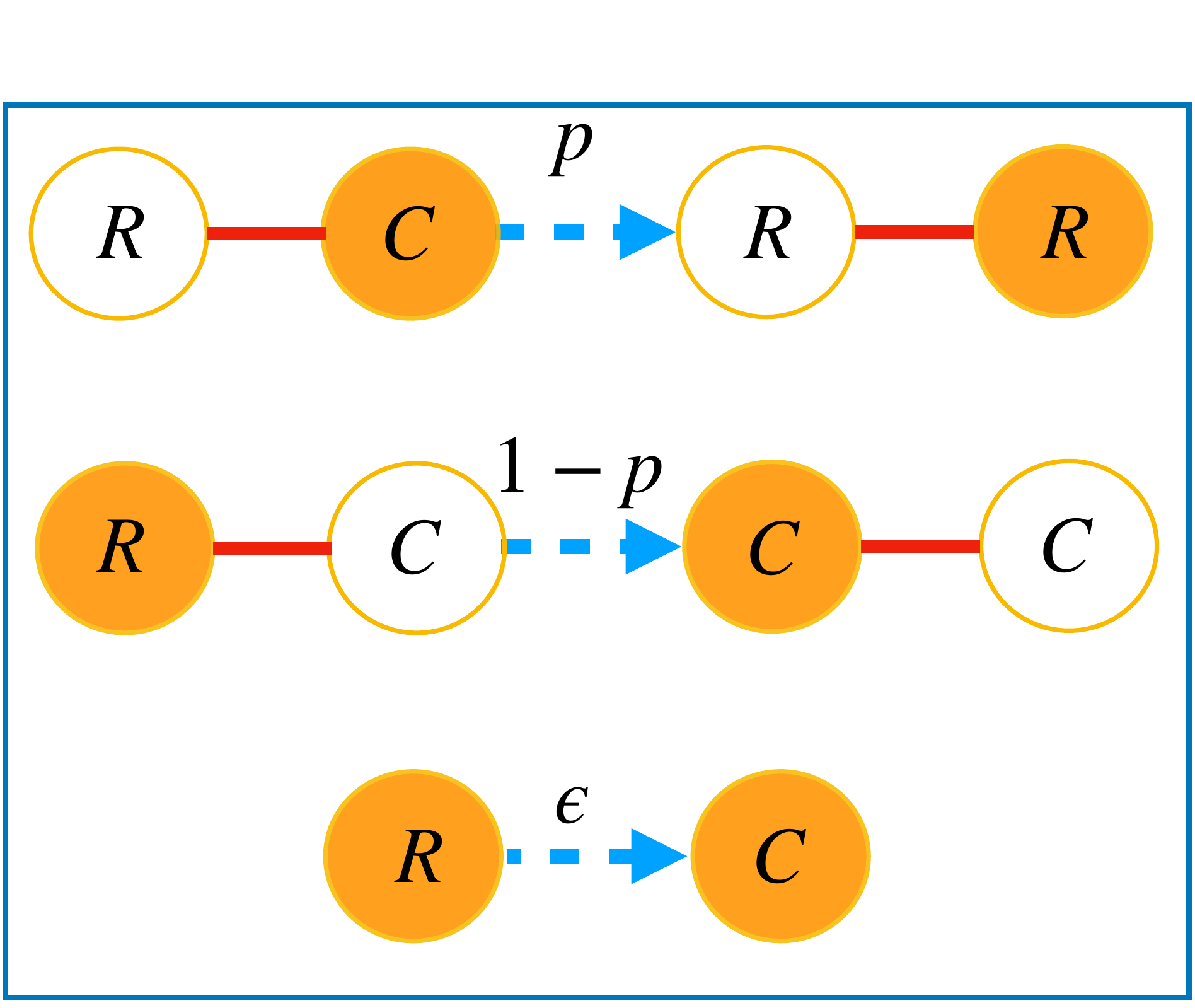} 
\caption{Illustration of the 3-state  I-voter dynamics. Circles with the legend $C$, $L$, and $R$ denote the centrists, leftists, and rightists, respectively. Lines indicate the interactions between two connected agents, while dashed arrows depict how the \emph{highlighted} agent changes his/her opinion upon interactions. The updates that are independent of the agent interactions include the decay of leftist (rightist) to centrist.
The parameters $p$, $1-p$, and $\epsilon$ are the respective rates of opinion updates.
} 
    \label{fig:fig1}
\end{figure}

To implement the opinion formation process, we employ asynchronous updating, in which each agent is assigned its own independent Poisson clock, all with the same unit rate. If it is in the state $0$, then when
its Poisson clock rings it changes its state from $0$ to $1(-1)$ with probability $p$ if the state of a randomly selected neighbour is right (left); otherwise, it remains unchanged. Similarly, if it is in the state $1(-1)$, it changes its
state to $0$ with probability $\epsilon$,  regardless of its neighbors' state, and with probability $1-p$ if the state of a randomly selected neighbour is center. We simulate this model by the Gillespie algorithm \cite{GILLESPIE}.

\section{Results}
\label{sec:results} 

\subsection{The steady-state fraction of centrists} Let $\rho_+$,  $\rho_-$ and $\rho_0$ denote the densities of rightists, leftists and centrists, respectively. In the Appendix \ref{sec:Derivation1}, we show that, for a \emph{fully-connected} network and in the limit of \emph{infinite} system-size $N\rightarrow \infty$, if $a := 2p - \epsilon -1 >0$, then the fraction $\rho_0$ of centrists is given by
\begin{equation}
    \rho^{*}_0 = \frac{\epsilon}{2p-1}
\label{invariance}
\end{equation}
We stress that a full parameter scanning over all combinations of $p$ and $\epsilon$ will result in 2 phases, $\rho^{*}_0 = \epsilon/(2p-1)$ for $a>0$ and $\rho^{*}_0 = 1$ (i.e., a society consisting of only centrists) for $a<0$. Since we are not interested in the latter phase without any extremists, throughout the paper, we only consider the case of $a>0$,  representing the case that the probability to move to the center is much smaller than the probability to become more extreme. 
 Next, for a system of finite size $N$, where finite-size fluctuations need to be taken into account,  we first represent the model as a chemical reaction network and then use a continuous-time Markov Chain to describe the evolution of the distribution of different opinions considered as chemical species.
Our approximation is based on a truncation of the moment hierarchy associated with this distribution up to the second order. This yields $\rho_0^* = \langle \rho_0\rangle_*$, where $\langle \cdot\rangle_*$ denotes averaging taken by the stationary distribution and with slight abuse of notation, $\rho_0$ denotes the fraction of centrists in a single realisation of the model dynamics. Using the same approximation scheme, in  the Appendix \ref{sec:Derivation2}, we also obtain the
variance of  $\rho_0$ in the steady state:
\begin{equation}
 {\rm Var}(\rho_0):= \langle  \rho_0^2 \rangle_* - \langle  \rho_0 \rangle_*^2  = \frac{\epsilon}{2(2p-1)}\frac{1}{N}
\label{variance_three_state}
\end{equation}
This shows how fluctuations due to finite-size alter the mean-field prediction. In particular,  either a high decay towards the neutral state or a low persuasion results in increased variance.

Figure 2  demonstrates typical random trajectories of the I-voter model. Here, in agreement with the mean-field prediction, we find $\langle \rho_0\rangle_* = 0.25$ for  $(p, \epsilon) = (0.7, 0.1)$. Note that because of the symmetry in the dynamical laws for leftists and rightists, we always obtain a statistical equality between the fraction of leftists and that of rightists if started from an unbiased initial condition with the same number.  This is observed in figure 2  for $\rho_+$ and $\rho_-$. 
In addition, we also verify Eq. \eqref{variance_three_state}, where fluctuations are observed to be within the shaded area bounded by two bands $\langle \rho_0\rangle_* \pm {\rm std}(\rho_0)$, i.e. within one standard deviation of the mean-field solution Eq. \eqref{invariance}.

\begin{figure}
\centering
\includegraphics[width=.64\linewidth]{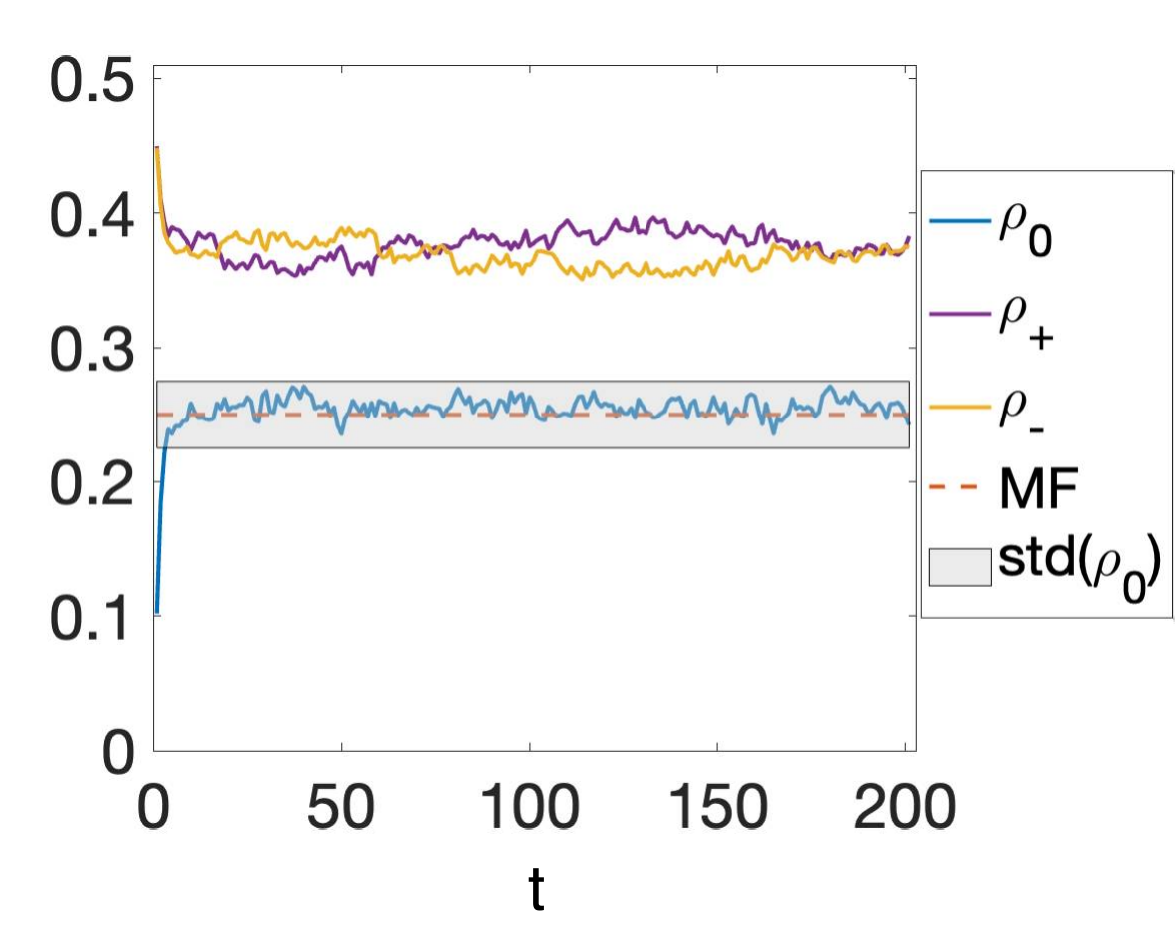} 
\caption{The density of opinions as function of time. ``MF" denotes the mean-field prediction $\rho_0^*$ and is depicted by the dashed red line. Stochastic trajectories are generated by the Gillespie algorithm for $N=100$, and then averaged over 100 independent runs on an all-to-all network for $\epsilon=0.1$ and  $p=0.7$. The fraction of centrists converges to $\rho^{*}_0 = \epsilon/(2p-1) = 0.25$. The fraction of centrists, rightists and leftists are denoted by $\rho_0$, $\rho_+$ and $\rho_-$ respectively. The shaded grey area depicts the standard deviation derived from the mean-field approximation of the full dynamics as given in  Eq. \eqref{variance_three_state}.}
    \label{fig:fig2}
\end{figure} 

\subsection{The role of network connectivity}
 In a social network $\mathcal{G}$  with adjacency matrix $A_{ij}\in \{0,1\}$, a pair of agents $i$ and $j$ are connected if $A_{ij}=1$, and they do not interact if $A_{ij}=0$. 
 Let $\mathcal{V}$ denote the set of nodes in  $\mathcal{G}$. For every node $i \in \mathcal{V}$, we consider its local neighborhood  
 $\partial_i := \{j \in \mathcal{V}: A_{ij} =1 \}$ consisting of  its  nearest neighbors only. A node $i$'s degree then is given by the number of its neighbors $\kappa_i:= \sum_{j\in \partial_i}  A_{ij}$. The level of connectedness in society is quantified by the average number of connections per node: $\kappa = N^{-1}\sum_i \kappa_i$. 
To study the effect of network connectivity on the opinion distribution, we consider $N$ agents, each has a probability of flipping its  opinion
depending on the states of its nearest neighbors in $\mathcal{G}$.

Let $\mathbb{P}(\mathbf{x},t)$ denote the joint distribution  to observe a  configuration  $\mathbf{x}:= (x_1, x_2,\cdots, x_N)$ at time $t$. The Appendix  \ref{sec:Derivation3}  provides details of how  this distribution  evolves according to a master equation Eq.  \eqref{full_master5}, whose transition rates $\mathbf{W}(\mathbf{x}'|\mathbf{x})$ 
from  $\mathbf{x}$ to  $\mathbf{x}'$ between $t$ and $t+ dt$ are given in  Eqs.  \eqref{full_rate_matrix}-\eqref{all_transitions2}. This master equation is not solvable in general, so to construct a mean-field  theory for our model, we introduce the averaged dynamical variable $\sigma_i(t)$ defined as   the probability that  node $i$ is \emph{not} a centrist at time $t$:
\begin{equation}  \sigma_i(t):= \sum_{\{\mathbf{x}\}} \mathbb{P}(\mathbf{x},t)\Big[\delta_{x_i,1}+\delta_{x_i,-1}\Big]
\label{local_activity}
\end{equation}
and the probability $\rho_i^{(0)}(t)$ that a node $i$ is a centrist at time $t$:
\begin{equation}
  \rho_i^{(0)}(t):=  \mathbb{E}\Big[\delta_{x_i,0} \Big] =  \sum_{\{\mathbf{x}\}} \mathbb{P}(\mathbf{x},t) \delta_{x_i,0}
= 1- \sigma_i(t) 
\label{full_master4} 
\end{equation}
where $\delta_{x,y}$ is Kronecker's delta and  the sum $\sum_{\{\mathbf{x}\}}$ is carried over the entire phase space of $3^N$ configurations.  
In  the Appendix \ref{sec:Derivation3}, we derive from  Eq. \eqref{full_master5} the following set of  $N$ \emph{approximate} mean-field equations for $\sigma_i$, which measure $i$'s averaged extremeness:
\begin{equation}
\frac{d\sigma_i}{dt} =  - \epsilon \sigma_i(t) +\frac{p \rho_{i}^{(0)}}{\kappa_i}\, \sum_{j\in \partial_i} \sigma_j(t)  - \frac{1-p}{\kappa_i}\sigma_i(t)\sum_{j\in \partial_i} \rho_{j}^{(0)}
\label{full_master3} 
\end{equation}

 To quantify the level of polarisation, we introduce the following measure:
\begin{equation}
\mathcal{P} = 1-\frac{1}{N}\sum_i\rho_i^{(0)}(t)  - \left(\frac{1}{N}\sum_i \mu_i\right)^2 
\label{polarisation}
\end{equation}
where
\begin{equation}  \mu_i(t):= \sum_{\{\mathbf{x}\}} \mathbb{P}(\mathbf{x},t)\Big[\delta_{x_i,1} -\delta_{x_i,-1}\Big]
\label{local_magnetisation}
\end{equation}
This measure  is  in line
with the idea that polarized societies typically lack
 a neutral attitude as common ground
for global consensus and have a high variance of opinions \cite{Bramson}. If the probability of being centrist for any individual is low  (for instance, a small fraction of respondents who chose the middle category in an opinion poll), and it is equally likely to be either left or right, $\mathcal{P}$ will have high value. So $\mathcal{P} \in [0,1]$, $\mathcal{P}=0$ means no polarisation and $\mathcal{P}=1$ indicates the highest polarisation level -- this latter case corresponds to a population containing, on average, as many rightists as leftists. We remark that  $\mathcal{P}$ can be considered as a discrete version of the well-established  \emph{polarization index} introduced in \cite{Morales} when the distribution of positive and that of negative opinions collapse into Dirac delta functions centered at $1$ and $-1$, respectively.

Real social networks are often small-worlds and  structurally heterogeneous with an abundance of well-connected nodes (hubs) \cite{WattsStrogatz, WILLIAMS2015, Iniguez2023, Borge-Holthoefer, Zhao2025}. To capture the effect of such heterogeneity, in Fig. \ref{fig:fig41} \textbf{(a)} we fix $p=0.7$, $\epsilon = 0.1$ and investigate  scale-free networks  of $N=2000$ with the exponent $2.8$ at varying mean degree $\kappa$  generated by the Chung–Lu random graph model \cite{Fasino}. In Fig. \ref{fig:fig41} \textbf{(b)} we next
test if the particular network structure has an influence
on the results by fixing  $\kappa=50$ and comparing scale-free with small-world   topology gererated using
  the Watts-Strogatz parameter $r$ --  the probability to reconnect a link from any node to any other node in the network \cite{WattsStrogatz} [$r= 0$
corresponds to   ring networks, $r =1$ --  Erdős–Rényi 
graphs].  Apart from the speed of convergence, there is no marked difference in terms of the steady-state behavior between these different topologies. This steady-state can be decently predicted by our MF approximation in Eq. \eqref{full_master3} as shown by the dashed line in Fig. \ref{fig:fig41} \textbf{(b)}. We note that the mean-field solutions can depend strongly on the initial conditions. We  discuss this point in details in the Appendix \ref{sec:Derivation3}.

\begin{figure}
\centering
\includegraphics[width=.48\linewidth]{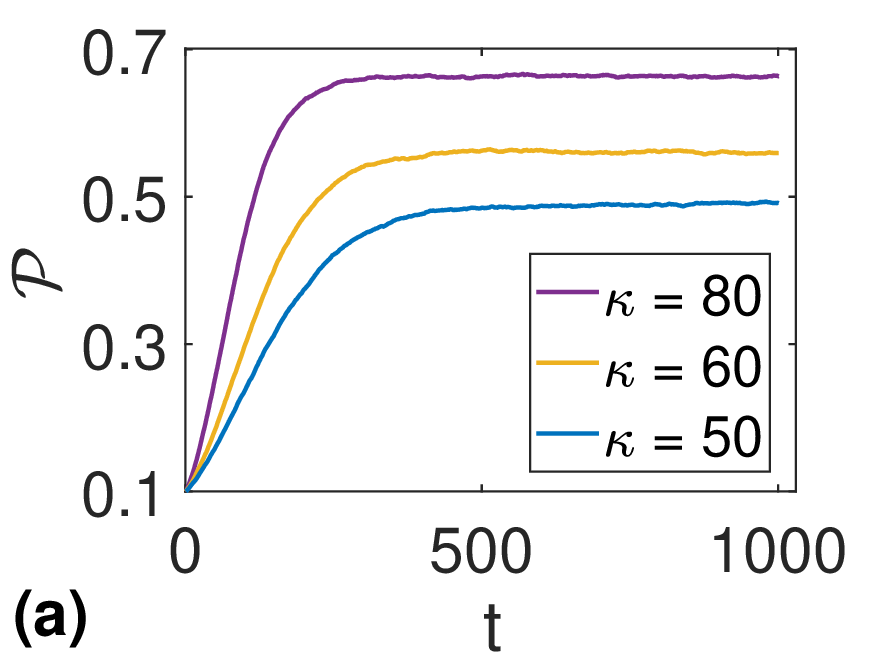}
\includegraphics[width=.48\linewidth]{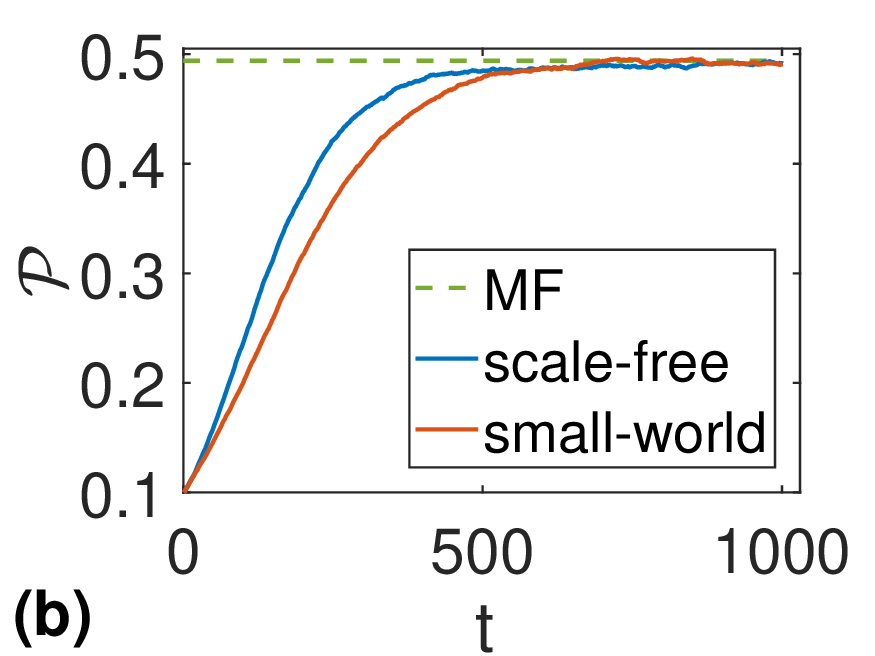}
\caption{The polarisation measure $\mathcal{P}$ in scale-free graphs with  the exponent $2.8$ and  various mean degree $\kappa= 50,60,80$ \textbf{(a)}.  $\mathcal{P}$ in
social networks with fixed $\kappa =50$ but different topologies \textbf{(b)}. Continuous lines are stochastic trajectories
generated by the Gillespie algorithm for $N=2000$ and then averaged over 100 independent
runs at fixed $\epsilon=0.1$ and $p=0.7$. The initial fractions of leftists and rightists are equal to $0.05$ in all these runs. Dashed line depicts the steady-state value obtained from the ``MF" solutions of Eqs.  \eqref{full_master4} and \eqref{full_master3} with suitable initial conditions.}
    \label{fig:fig41}
\end{figure}

In Figure \ref{fig:fig3}, we compare our mean-field predictions with simulations on  networks of varying average degrees $\kappa$.  We obtain good agreement for dense networks (i.e., $\kappa = O(N)$), but deviations as the network becomes sparser. This can already be observed for $\kappa =60$.
Overall, both simulations and mean-field predictions show that as $\kappa$ increases,  $\mathcal{P}$ increases, indicating that polarization level rises with increasing connectivity. To check whether this behaviour remains robust with variations in $p$
and $\epsilon$, provided that $a= 2p-\epsilon-1>0$, we compute  the phase diagram of $\mathcal{P}$ in Fig. \ref{fig:fig4}.  We propose to use the ratio $\epsilon/(2p-1)$ as an effective parameter controlling  the level of involvement that  intuitively decreases with increasing $\epsilon/(2p-1)$. Polarisation is more likely to occur in a society with highly involved agents: $\mathcal{P}$ vanishes as this ratio increases beyond a critical value and the faster decay of the individual involvement, the lower $\mathcal{P}$ is. Apart from the special case of $\epsilon \rightarrow 0$, for a given level of involvement, a high level of polarisation $\mathcal{P}\simeq 1$ can only be achieved at sufficiently large degree $\kappa$. We note that our approximation qualitatively reproduces the boundary between polarised and non-polarised phases, but it becomes more inaccurate as $\epsilon/(2p-1)$ increases. We remark that our results remain robust wrt the inclusion of noise as shown in  the Appendix  \ref{sec:noise}.

\begin{figure}
\centering
\includegraphics[width=.75\linewidth]{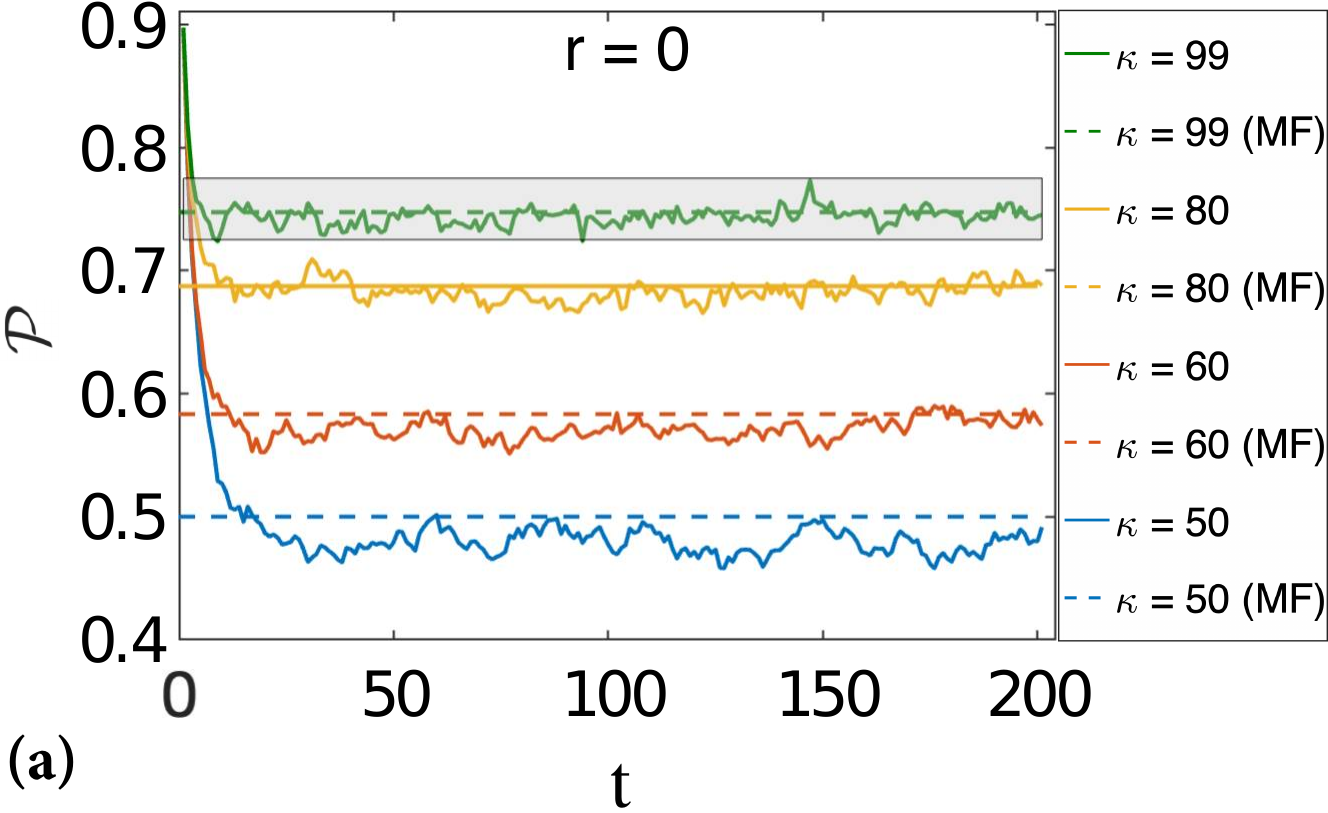} 
\includegraphics[width=.75\linewidth]{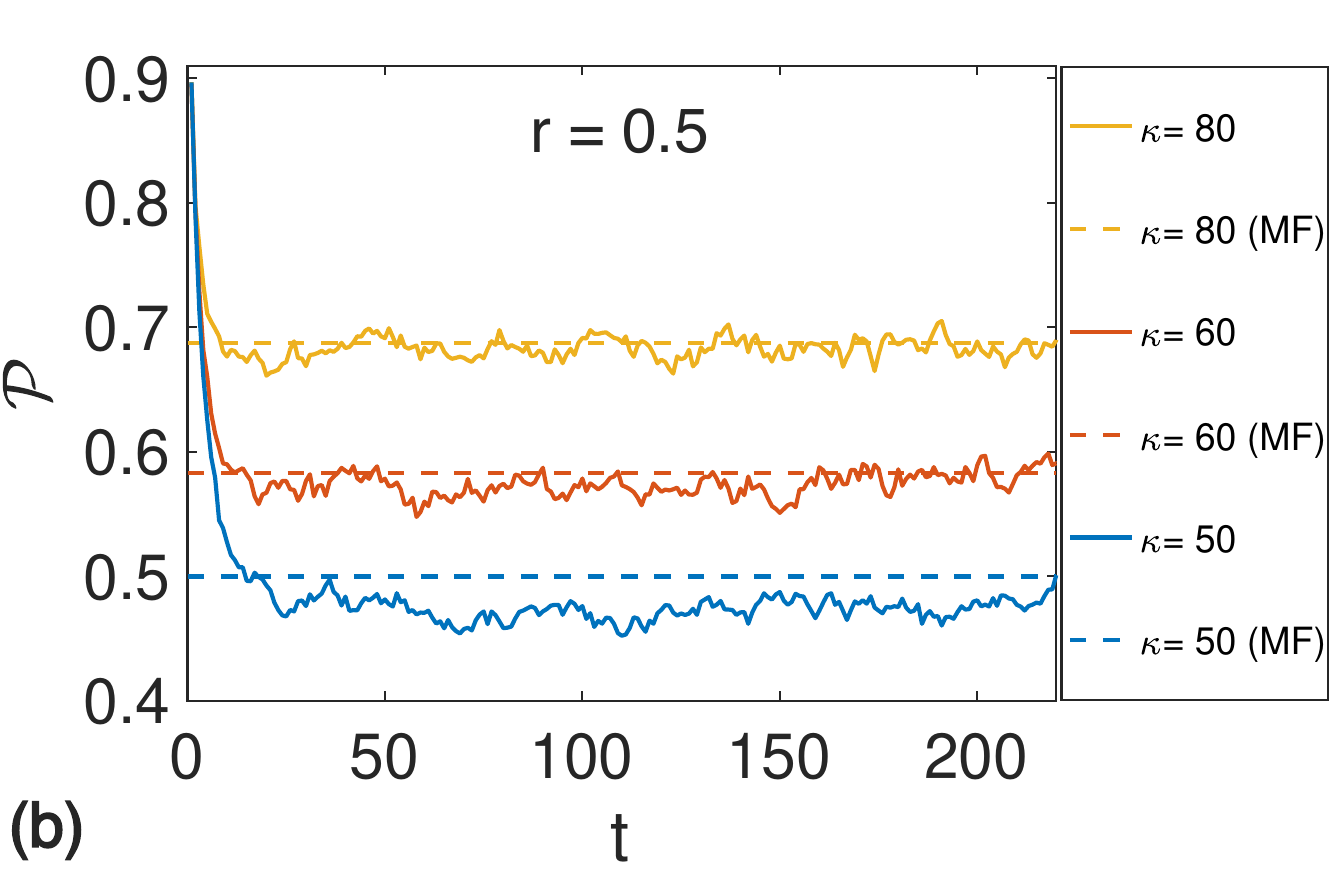}
\includegraphics[width=.75\linewidth]{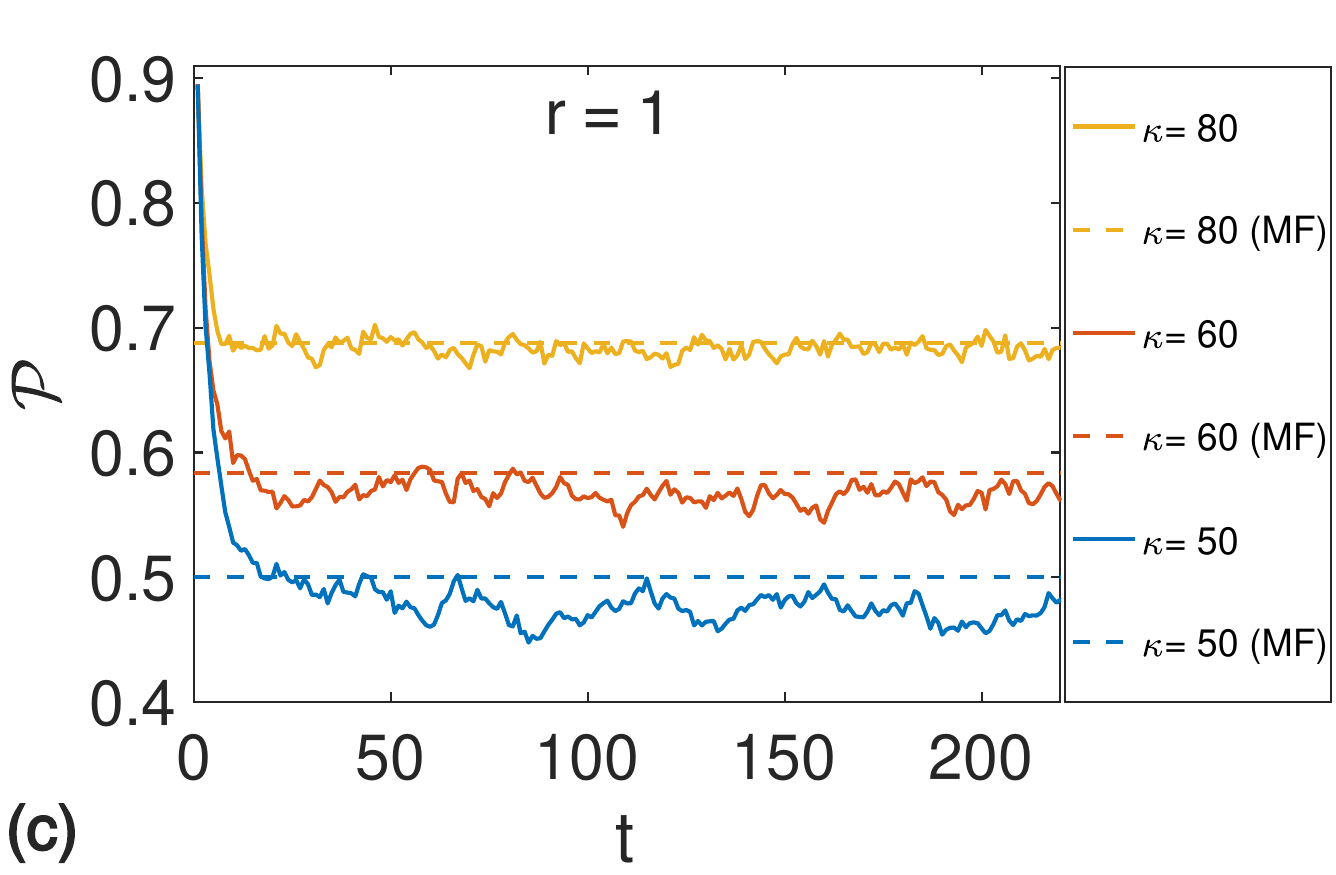}
\caption{The polarisation measure $\mathcal{P}$ for a social network with a ring topology \textbf{(a)} ($r= 0$); Watts-Strogatz small-world network \textbf{(b)} ($r= 0.5$) and Erdős–Rényi random graph  \textbf{(c)} ($r= 1$), where $r$ is the rewiring probability \cite{WattsStrogatz}, all with various degrees $\kappa$. $\mathcal{P}$  increases with increasing $\kappa$, showing polarisation level rises up in more connected social networks. Dashed lines depict the ``MF" prediction according to Eqs.  \eqref{full_master4} and \eqref{full_master3}. Continuous lines are stochastic trajectories generated by the Gillespie algorithm for $N=100$, and then averaged over 100 independent runs. The shaded grey area depicts the standard deviation derived from the mean-field approximation of the full dynamics as given in Eq. \eqref{variance_three_state}.  Here  $\epsilon=0.1$, $p=0.7$; the initial fractions of leftists and rightists are equal $0.45$.}
    \label{fig:fig3}
\end{figure}

\begin{figure}
\centering
\includegraphics[width=.48\linewidth]{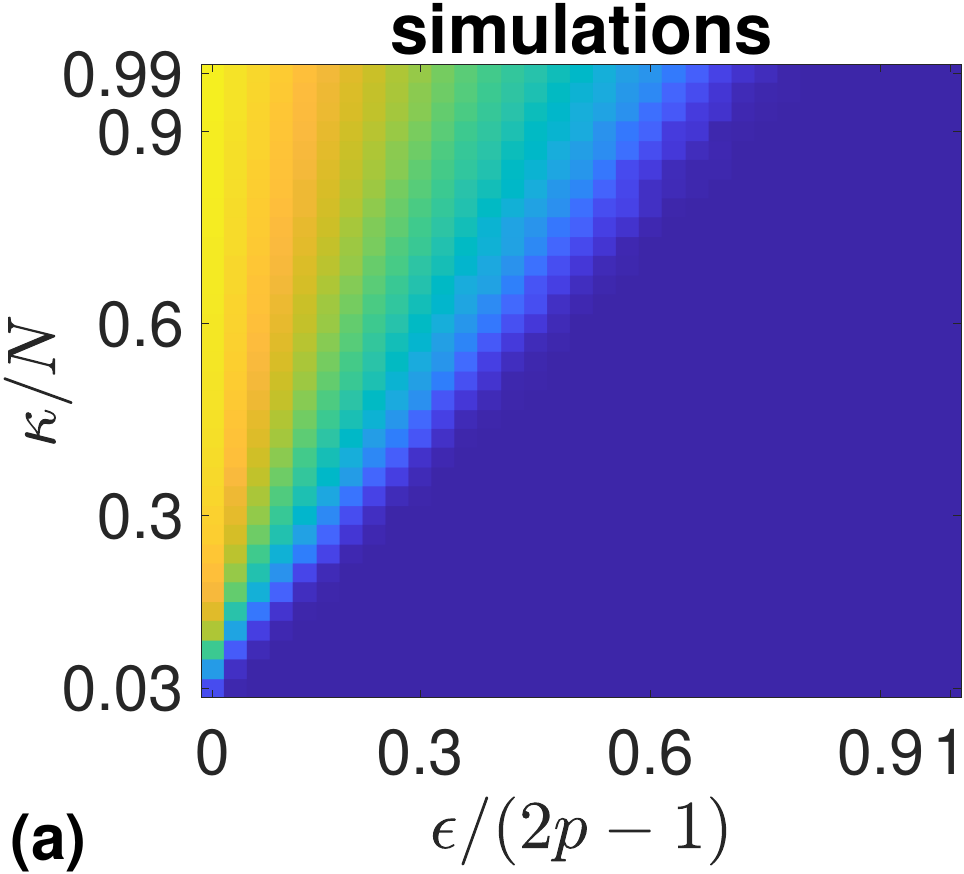}
\includegraphics[width=.48\linewidth]{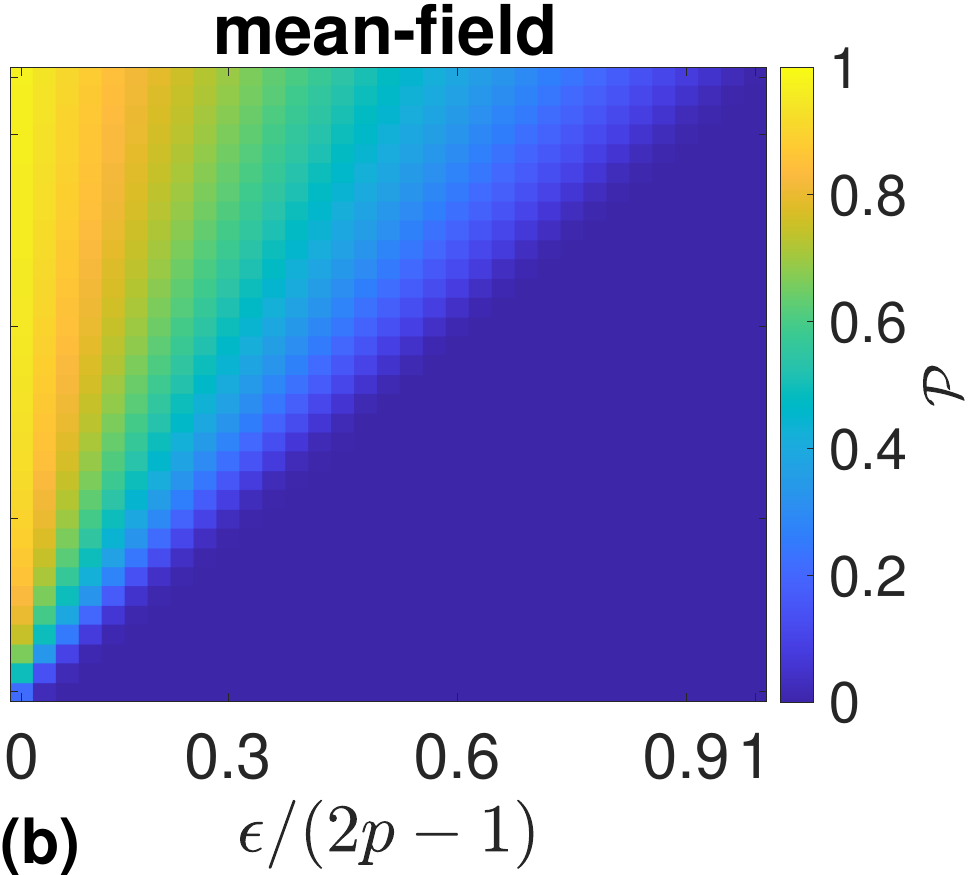}
\caption{The polarisation measure $\mathcal{P}$ (coded by color: yellow for $\mathcal{P} =1$ and dark blue for $\mathcal{P} =0$   ) for a social network with a ring topology and various degrees computed from simulations \textbf{(a)} and from MF  solution to Eqs.  \eqref{full_master4} and \eqref{full_master3} \textbf{(b)}.  In both panels, $\mathcal{P}$ is shown as function of $\kappa/N$-- the average degree scaled by the system size on y-axis and $\epsilon/(2p-1)$--the effective parameter controlling the level of involvement on x-axis. We fixed $p=0.7$ and increase $\epsilon$, while keeping  $\epsilon/(2p-1) \in [0,1]$. The level of involvement decreases as this ratio increases. Here $N=100$ and the initial fractions of leftists and rightists are equal to $0.45$ in Gillespie simulations.}
    \label{fig:fig4}
\end{figure}

\subsection{n-state model}
A natural extension of the 3-state I-voter model is the one that includes two extra states $x_i=+2$ and $x_i=-2$ that we call the 5-state I-voter model.  Here (i) decay means that an agent moves to an opinion state that is one level less extreme with probability (per unit time) $\epsilon$;  (ii) persuasion can happen only when $\big|x_i-x_j\big| = 1$ so that (without loss of generality, we consider $|x_i|<|x_j|$) either $x_i$ goes one-level more extreme with probability  $p$ or   $x_j$  goes one-level less extreme with probability $1-p$;  and (iii) the reinforcement of extreme opinions can only occur between similar agents following their interaction so that if $x_i = x_j = \pm 1$, then both become one-level more extreme with probability $\gamma$.  The parameter $\gamma$ describes  an increased likelihood of moving towards  more extreme opinions when individuals engage in  discussion with like-minded others.  This phenomenon is known as group polarization \cite{LEE, Strandberg,HOBOLT, Zheng, Kuhn}. For example, in the so-called French Jury Study \cite{Moscovici1969}, French participants who already had a favorable attitude toward then-President Charles de Gaulle were asked to discuss their opinions in small groups. After the group discussion, their positive opinions became even more positive. Similarly, participants who disliked American foreign policy became even more negative about it after discussing it with like-minded others. Other evidence of this mechanism has been reported recently in  online platforms, such as Reddit and Gab \cite{Cinelli}. Therefore, we note that the implementation of the $\gamma$-based mechanism requires a two-body interaction, whereas that based on $\epsilon$  is a one-body effect. As a result, the effectiveness of the former is determined by the mean number of connections $\kappa$, while the latter is independent of $\kappa$.  Adding pairs of states $x_i=\pm 3$, $x_i=\pm 4$, $\cdots$, while using the same rules for the 5-state model, results in the 7-state, 9-state models and so on.  Figure 6 \textbf{(a)} illustrates the 5-state I-voter model with $x_i=-2$ denoted by $L_2$ and  $x_i=2$ -- by $R_2$.

In Figure 6 \textbf{(b)} we observe that while the steady-state fraction of centrist is invariant wrt the introduction of $\gamma$ and two extra states, the underlying dynamics change in comparison to the 3-state I-voter model as shown in the inset. Here, $\rho_+$ and $\rho_-$, both relax to values close to zero (but strictly positive as long as $\epsilon>0$), while the densities of $R_2$ and $L_2$, denoted $\rho_{2+}$ and $\rho_{2-}$, respectively, reach significantly higher values, indicating the emergence of more extreme opinions under the strong influence of $\gamma $.
 In   the Appendix  \ref{sec:Derivation4} we derive the independence of $\rho^*_0$ on $\gamma$ within the mean-field description as well as by truncating at the second order in the moment hierarchy. Next, we generalise the use of the polarisation measure  $\mathcal{P}$ proposed in Eq. \eqref{polarisation} to the  $n$-state model.
To this end, we modify the expressions for  $\sigma_i, \mu_i, $ and $\rho_i^{(0)}$ as follows:
\begin{equation} 
\begin{aligned}
\sigma_i(t) &:= \sum_{\{\mathbf{x}\}} \mathbb{P}(\mathbf{x},t)\left\{\delta_{\displaystyle x_i,|x_i|}+\delta_{\displaystyle x_i,-|x_i|}\right\}\\ \mu_i(t) &:= \sum_{\{\mathbf{x}\}} \mathbb{P}(\mathbf{x},t)\left\{\delta_{\displaystyle x_i,|x_i|}-\delta_{\displaystyle x_i,-|x_i|}\right\} \\
\rho_i^{(0)}(t) &:=  \mathbb{E}\Big[\delta_{x_i,0} \Big] =  \sum_{\{\mathbf{x}\}} \mathbb{P}(\mathbf{x},t) \delta_{x_i,0}
= 1- \sigma_i(t)  
\end{aligned}
\label{modified_extremeness}
\end{equation}
 
In Figure 6 \textbf{(c)} 
we confirm  a similar increase of $\mathcal{P}$ with increasing $\kappa$ in this case.  Given that the measure $\mathcal{P}$ depends on the joint distribution of all agents across 5 distinct states, it is non-trivial to see how an invariant fraction of centrists alone can lead to the same increase of polarisation with the average degree $\kappa$. For now, we only remark that at the mean-field level, $\rho_0^*$ can be shown to be independent of $\gamma$ for any $n$-state I-voter. This suggests that our result on polarisation is general and is expected to go well beyond the 3-state and 5-state cases.

\begin{figure}
\centering
\includegraphics[width=.45\linewidth]{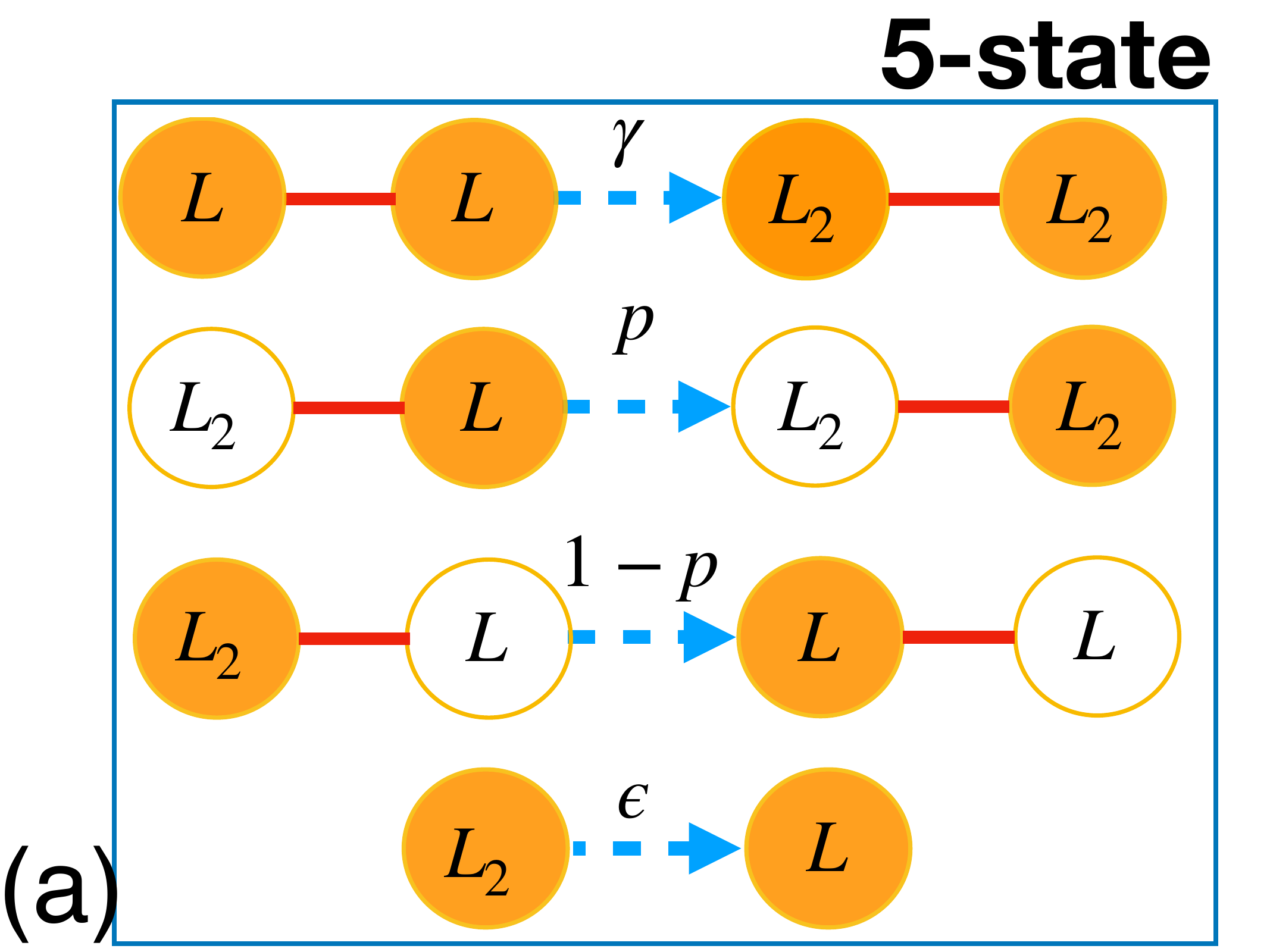} 
\includegraphics[width=.42\linewidth]{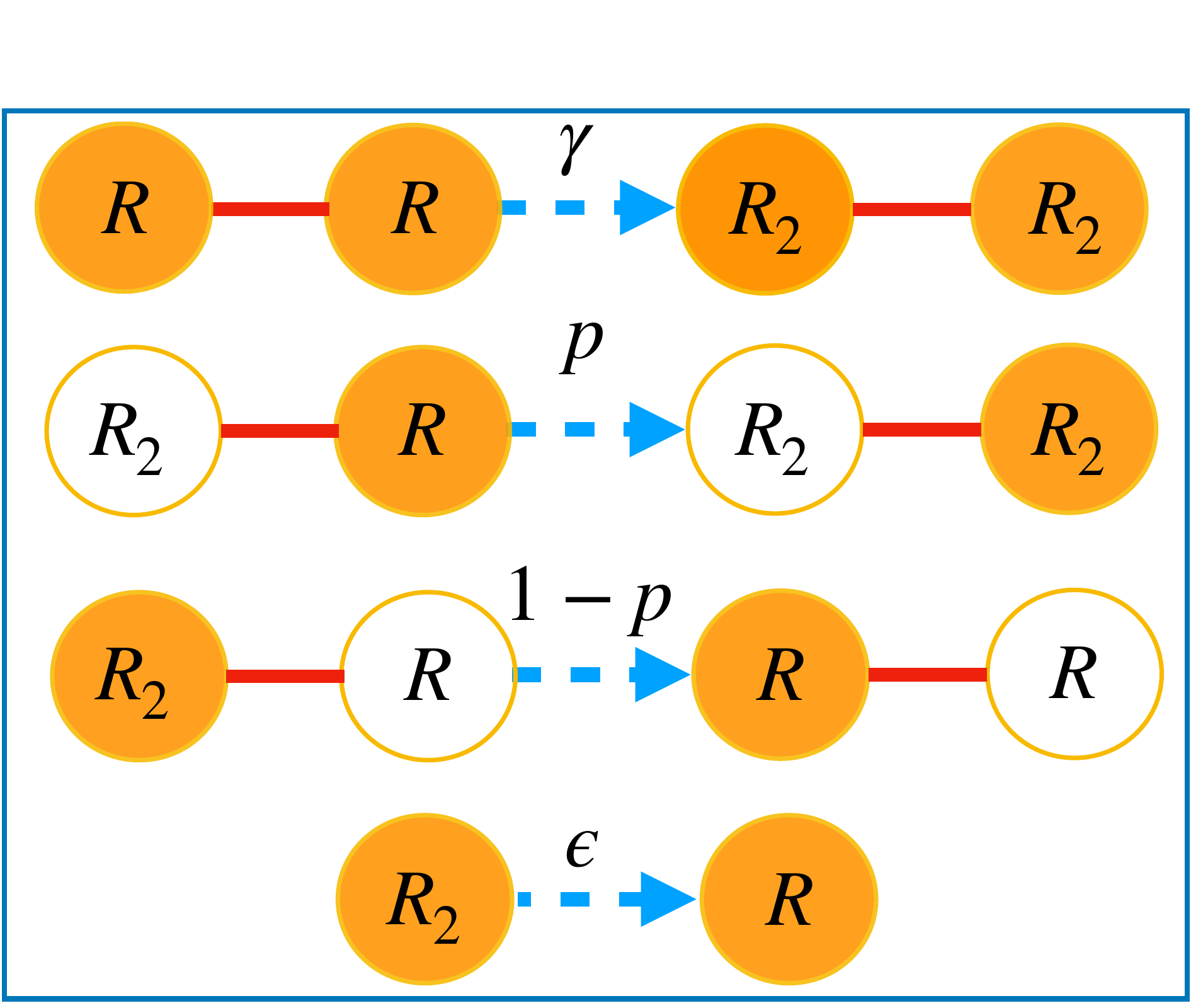}

\includegraphics[width=.75\linewidth]{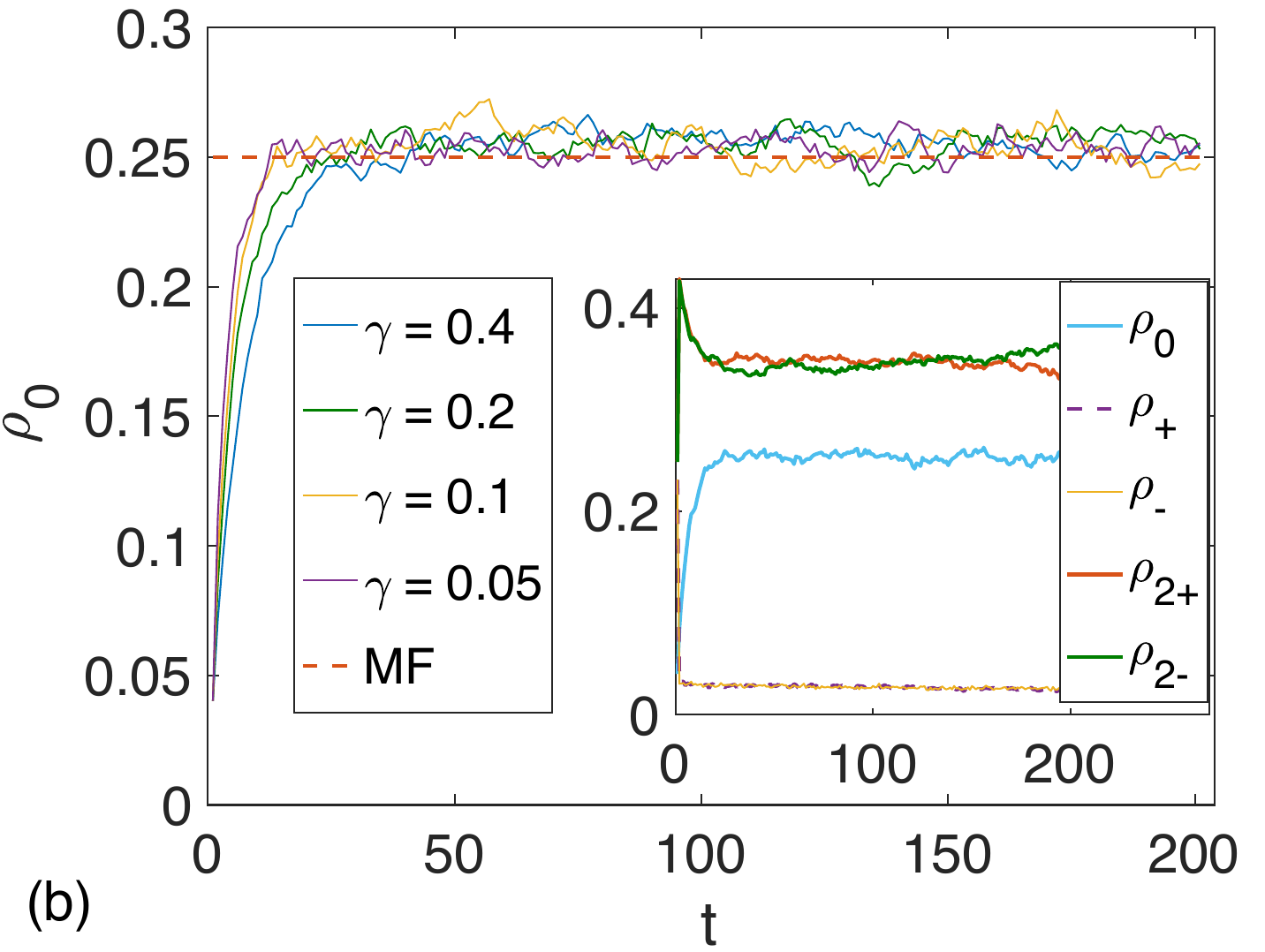} 
\includegraphics[width=.75\linewidth]{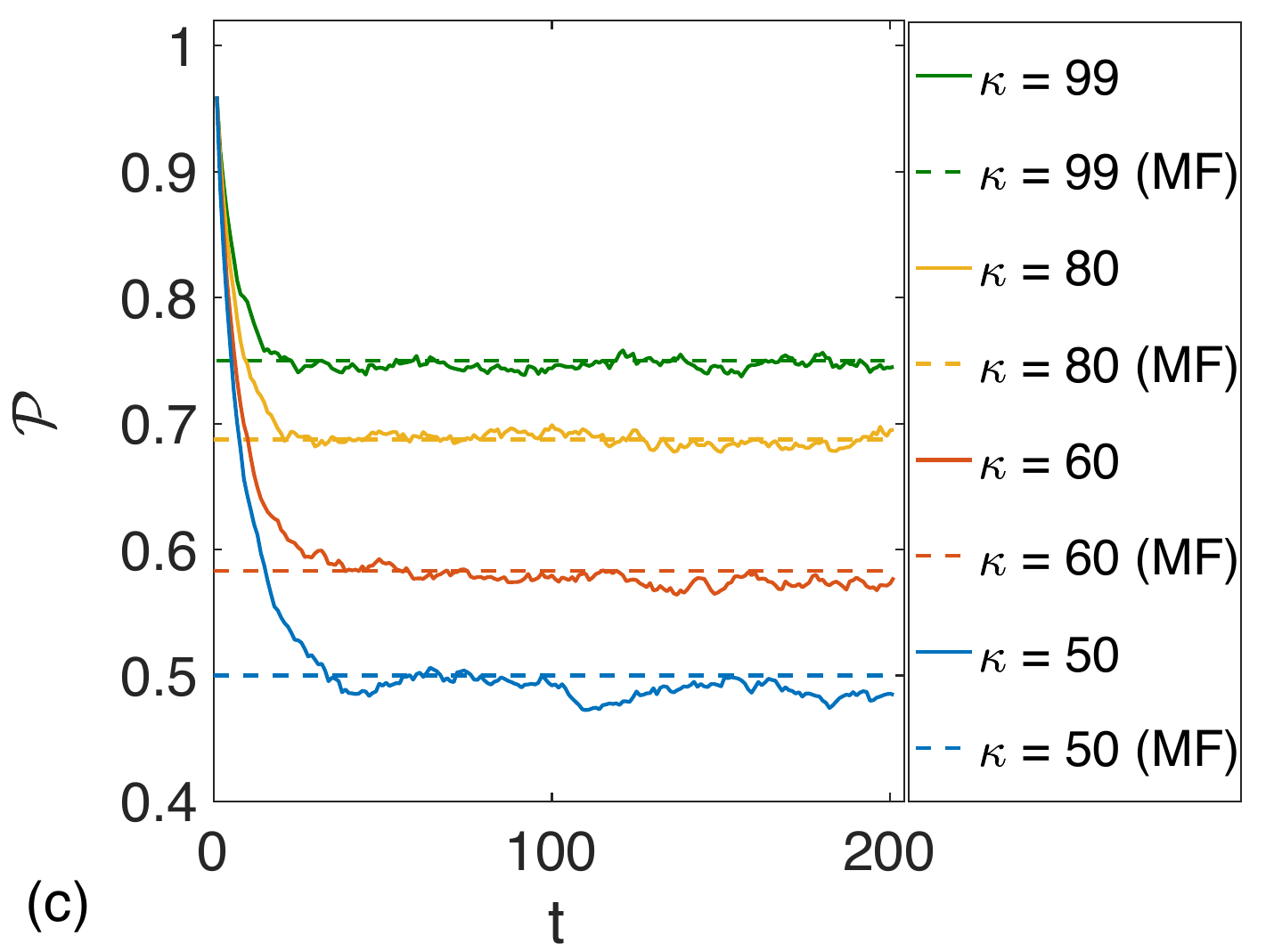} 
\caption{\textbf{(a)} Illustration of the 5-state  I-voter dynamics. In addition to the mechanisms plotted in Figure 1,  there are 8 extra. Circles with the legend $L_2$ and $R_2$ denote the state with $x_i=-2$ and $x_i =2$, respectively. Lines indicate the interactions between two connected agents, while dashed  arrows depict how the \emph{highlighted} agent changes his/her opinion upon interactions. The updates that are independent of the agent interactions include the decay of an $L_2$ ($R_2$) agent to leftist(rightist). \textbf{(b)} Main: the fraction of centrists $\rho_0$ in the 5-state  model in an all-to-all graph for  $\gamma = 0.05, 0.1, 0.2, 0.4$, where ``MF" denotes the mean-field prediction $\rho^{*}_0 = \epsilon/(2p-1) = 0.25$ and is depicted by the dashed red line;     Inset: the density of different opinions for $\gamma=0.2$.  The fraction of  rightists (leftists)   and that of $x_i=+2$ ($x_i=-2$) are denoted by $\rho_+$($\rho_-$) and $\rho_{2+}$ ($\rho_{2-}$) respectively. \textbf{(c)} The polarisation measure $\mathcal{P}$ of 5-state model as defined in  Eq. \eqref{polarisation} but with $(\rho^{(0)}_i,\mu_i)$ given in Eq. \eqref{modified_extremeness},  for  varying degrees $\kappa$  with fixed $\gamma=0.2$.
In \textbf{(b)} and \textbf{(c)}, stochastic trajectories are generated by the Gillespie algorithm for $N=100$, and then averaged over 100 independent runs
for $\epsilon=0.1$,  $p=0.7$.}
    \label{fig:fig5}
\end{figure} 
\section{Discussion}
\label{sec:discussion}
Modern societies are characterized by unprecedented interconnectedness due to technological advancements, particularly social media platforms. These platforms facilitate rapid and widespread dissemination of information and opinions, significantly altering the dynamics of social interactions. In this context, the level of individual engagement, or involvement, becomes a critical factor in understanding how opinions are formed and sustained. In this regard,    we proposed  a new way to understand the rise  of opinion polarisation in increasingly connected societies as a consequence of the joint effect of involvement characterised by $(p, \epsilon)$ and the mean degree $\kappa$.

We found that, for fixed values of $p$ and $\epsilon$, denser networks exhibit higher levels of polarization.   This is shown to be the case in both the 3-state and 5-state I-voter models but is expected to hold for $n$-state dynamics with $\gamma>0$ capturing a tendency of extremists to become even more extreme after discussion with
like-minded others. These results are in qualitative agreement with recent empirical findings \cite{Smith2024, Kazmina}. A consequence of these findings is that an increase in social relations, either in person or virtual, may lead to polarisation while a decrease in social relations may lead to depolarisation. 
For future work, we would investigate intervention strategies to shift the system between polarized and neutral states. Reducing $p$ and increasing $\epsilon$ can decrease polarization. Centrists should be more resistant to extremist arguments, and involvement of extremisms should diminish more rapidly.

We note the following limitations. While the assumptions and predictions of our model align with a significant portion of the empirical literature (see main text for references), it does not yet offer quantitative predictions. A first step would be to estimate model parameters from a real dataset \cite{GALESIC2019275, Vendeville, Peralta}.
For the sake of analytical treatment, we have studied only homogeneous populations of agents with the same parameters $p$ and $\epsilon$, neglecting possible important effects of heterogeneity in the model parameters. A natural step then is to consider the case where each agent is characterized by individual values of $p_i$ and  $\epsilon_i$. In this case, there might exist multiple stable steady states induced by individual heterogeneity. For this, one would compute the mean first passage (convergence) time to reach a given steady state and the attractor-switching time. 

Next, it is worth exploring the effect of antagonistic ties \cite{castro2025}, which have been shown to play an important role in mitigating ideal polarization within village networks  \cite{Ghasemiana}. A reduction of opinion polarization by
incidental similarities, i.e. shared personal traits between those individuals who hold different opinions on a political issue, has recently been found in \cite{Balietti}. Therefore, it would be interesting to include demographic and biographical features, such as age, gender, language, nationality, and personal interests into our model and study how these features affect the ideological dimension. This will facilitate comparisons with the large-scale experiment of \cite{Balietti} and the Axelrod model's prediction \cite{RAxelrod1997}.

 
\begin{acknowledgments}
 We thank Ben Meylahn and Wout
Merbis for helpful comments. This work was supported in part by the Dutch Institute for Emergent Phenomena (DIEP) cluster at the
University of Amsterdam under the Research Priority Area \emph{Emergent Phenomena in Society: Polarisation, Segregation and Inequality} and the programme Foundations and
Applications of Emergence (FAEME). 
\end{acknowledgments}


\appendix

\section{Note on related models}
\label{related_model}
 The hierarchical Ising opinion model (HIOM) \cite{HIOM}:  The HIOM \cite{HIOM} is a complex cascading transition model that captures the interplay between individual dynamics and polarization across individuals. The HIOM conceptualises an agent’s individual attitude as a network of beliefs, feelings, and behaviours towards an issue \cite{Dalege2018,Dalege2024}. The alignment of nodes in an individual's attitude network depends on involvement. In lowly involved agents attitudes are weak and inconsistent, while highly involved agents develop extreme opinions. Changes in information (the external field) can lead to sudden jumps and hysteresis. In the HIOM
involvement plays a double role. First, agents with high involvement initiate more interactions and are more persuasive than less-involved ones. Second, involvement generally decays but increases due to  interactions. Therefore, similar to the I-voter model, polarization increases in highly connected societies. However, due to the complexity of the setup, an analytical treatment of this effect is not feasible.

The constrained 3-state voter model on all-to-all graphs \cite{Vazquez2004} features  a steady state, in which  either  no neutral opinion exists or  a consensus on only one of the three opinions is reached. 
This means that the first kind of steady states of this model
can be considered as the  $\epsilon\rightarrow 0^+$ limit  of the  I-voter dynamics which also relaxes to a stationary mixture of  leftists and rightists, but without any centrists. However, due to the decaying effect  of  involvement that turns extreme agents to neutral ones at a rate $\epsilon> 0$, configurations in the I-voter model always include some fraction of centrists, making it different from the constrained 3-state voter model even in the mean-field limit.   Another  variant of the constrained voter model \cite{Mobilia2013} features a ``multi-opinion'' phase in the mean-field limit similar to ours, but this phase does not persist in finite populations due to demographic fluctuations.

\section{Derivation of Eq. \eqref{invariance}}
\label{sec:Derivation1}
The I-voter model with three states is described by  a set of two  ODEs for $\rho_-$ and $\rho_+$ (as $\rho_+ + \rho_- +\rho_0 = 1$) according to mass-action kinetics:    \begin{equation}
        \left \{ \begin{array}{l} \displaystyle 
\dot{\rho}_{+} = (2p-1)\rho_{+}\rho_{0} - \epsilon\rho_{+} \vspace{3pt}\\ 
\displaystyle  \dot{\rho}_{-} =  (2p-1)\rho_{-}\rho_0 -\epsilon\rho_{-}
\end{array} \right. 
\label{mean_field_limit}
    \end{equation}
By introducing $y= \rho_+ +\rho_-$ and $a= 2p-1 -\epsilon$, we get 
\begin{equation}
       \dot{y} =a y \left(1- \frac{y}{K}\right)\,,\qquad   K = \frac{2p-1 - \epsilon}{2p-1}  
\label{logistic}
    \end{equation}
  This takes the same form as the logistic equation that describes the growth of a species  with density $y(t)$
at a rate $ay$ and  decay  $ay^2/K$ with 
 the \emph{rescaled} carrying capacity in a given neighborhood $K<1$.
When $a>0$, the stable fixed point of the above dynamics is $y_* = K$. Since we are only interested in physical solutions with positive value, we consider only those pairs of $(p, \epsilon)$ that satisfy $2p-1>\epsilon$. From the conservation law $\rho^{*}_0+y_*=1$, we obtain
    $$ \rho^{*}_0 = \frac{\epsilon}{2p-1} $$
\section{Derivation of Eq. \eqref{variance_three_state}}
\label{sec:Derivation2} 
 We remark that by mapping the dynamical rules of I-voter updates onto a chemical reaction network scheme with three chemical species $L$ (leftist), $R$ (rightist) and $C$ (centrist),  both Eq. \eqref{mean_field_limit}  can be derived as the $N\rightarrow \infty$ -limit of the underlying master equation describing  the evolution of the reactant concentrations. The set of reactions for the model reads
\begin{equation}
        \left \{ \begin{array}{l} \displaystyle C +L \overset{k_1}{\longrightarrow} L+L
\vspace{2pt}\\ 
\displaystyle   L+C \overset{k_2}{\longrightarrow} C+C \vspace{2pt}\\ 
\displaystyle   L\overset{k_3}{\longrightarrow} C \vspace{2pt}\\ 
\displaystyle   C+R \overset{k_4}{\longrightarrow} R+R \vspace{2pt}\\ 
\displaystyle   R+C \overset{k_5}{\longrightarrow} C+C \vspace{2pt}\\ 
\displaystyle   R \overset{k_6}{\longrightarrow} C \vspace{2pt}\\ 
\end{array} \right. 
\label{baseline_reaction}
    \end{equation}
where $k_1=k_4=p$, $k_2=k_5=1-p$ and $k_3=k_6=\epsilon$. Our chemical reaction network formulation of the opinion dynamics is inspired by \cite{Armas_2025} and similar in spirit to  \cite{Holehouse2022}. 
We start by writing the quasi Hamiltonian  $H$ (i.e. the infinitesimal-time generator) for the master equation $\partial_t \mathcal{P} =H \mathcal{P}$, where for a \emph{discrete} probability distribution $w_{\mathbf{n}}(t) := w\big(n_L(t), n_R(t), n_C(t)\big)$ we introduce the associated generating function $\mathcal{P}(t, \mathbf{z}) = \sum_{\mathbf{n}} w_{\mathbf{n}}(t) z_L^{n_L} z_R^{n_R} z_C^{n_C}$, with the short-hand notations $\mathbf{n}:= (n_L, n_R, n_C)$ and $\mathbf{z}:= (z_L, z_R, z_C)$. Following \cite{baez2018quantum}, this Hamiltonian reads
\begin{equation}
\begin{aligned}
    H =&~ k_2\left\{\Big[\big( a_C^\dag \big)^2 - a_L^\dag a_C^\dag  \Big]a_L a_C  +\Big[\big( a_C^\dag \big)^2 - a_R^\dag a_C^\dag  \Big]a_R a_C\right\}\\ &+ p\Big[\big( a_L^\dag \big)^2 - a_L^\dag a_C^\dag  \Big]a_L a_C  + p\Big[\big( a_R^\dag \big)^2 - a_R^\dag a_C^\dag  \Big]a_R a_C \\ &+ \epsilon \Big[ a_C^\dag - a_L^\dag   \Big]a_L + \epsilon \Big[ a_C^\dag - a_R^\dag   \Big]a_R
    \end{aligned}
\end{equation}
where we have introduced the creation and annihilation operators for the leftists $a_L^\dag$ and $a_L$, as well as their counterparts $a_R^\dag$ ($a_C^\dag$)  and $a_R$ ($a_C$) for the rightists (centrists). 
Now let's introduce the number operators $\hat{N}_L =a_L^\dag a_L $, $\hat{N}_R = a_R^\dag a_R $ and $\hat{N}_C =a_C^\dag a_C $. Taking the derivatives of the generating functions we can evaluate the averaged number of leftists as follows:
\begin{equation}
    \frac{d}{dt}  \langle n_L\rangle = \frac{d}{dt}  \Big(\hat{N}_L\mathbb{P}\Big|_{\mathbf{z}=\mathbf{1}} \Big) 
\end{equation}
and similarly for the average number of rightists and centrists. 
The time-derivative of these averages  are  then given by
\begin{equation}
\begin{aligned}
    \frac{d}{dt}  \langle n_L\rangle &= 
  (2p-1)\langle  n_Ln_C \rangle -\epsilon \langle n_L\rangle \\   \frac{d}{dt}  \langle n_R\rangle  &= (2p-1)\langle  n_Rn_C \rangle -\epsilon \langle n_R\rangle
\\
 \frac{d}{dt} \langle n_C\rangle & = -(2p-1)\big\langle  (n_L+n_R)n_C \big\rangle +\epsilon \big\langle (n_L+n_R)\big\rangle 
\label{stochastic_number}
\end{aligned}
 \end{equation}
This set of unclosed equations is an example of the typical "moment closure" problem encountered in numerous fields \cite{Kuehn2016}, where we need to know $\langle  n_Ln_C \rangle$ for determining the evolution, for instance, of $\langle  n_L \rangle$. One can easily check that the second-order moments depend on the third-order moments, so on and so forth. For instance, we have for $\langle n_L n_C \rangle$ the following
\begin{equation}
 \begin{aligned}
    \frac{d}{dt}\langle n_L n_C \rangle =&~   (2p-1)\Big[\big\langle n_L n_C^2 \big\rangle -\big\langle n_L^2 n_C \big\rangle -\langle n_L n_R n_C\rangle\Big] \\ &+\epsilon \big\langle n_L(n_L-1) \big\rangle + \epsilon\langle n_L n_R \rangle -\epsilon\langle n_L n_C \rangle 
\label{second_moment}
\end{aligned}
\end{equation} 
Since the total number of agents $N=n_L +n_R +n_C$ is conserved in this case the last two terms can be expressed as
\begin{equation*}
    \begin{aligned}
    \epsilon\langle n_L n_R\rangle  &= N\epsilon\langle n_L\rangle - \epsilon\langle  n_Ln_C\rangle -  \epsilon \langle n_L^2\rangle   \\-\langle n_L n_R n_C\rangle & =  -N\langle n_L  n_C\rangle + \langle n_L n_C^2 \rangle +\langle n_L^2 n_C \rangle 
\end{aligned}
\end{equation*}
Substituting these expressions into Eq. \eqref{second_moment}, rescaling $p \rightarrow p/N$, $1-p \rightarrow (1-p)/N$ and introducing the  densities $\rho_+= n_L/N$, $\rho_-= n_R/N$ and $\rho_0= n_C/N$, we arrive at
\begin{equation}
\left \{ \begin{array}{l} \displaystyle 
\frac{d}{dt}\langle \rho_+ \rangle = (2p-1)\langle \rho_+ \rho_0 \rangle -\epsilon \langle \rho_+ \rangle \vspace{3pt}\\ 
\displaystyle  \frac{d}{dt}\langle \rho_+ \rho_0 \rangle = \Gamma\langle \rho_+ \rho_0 \rangle + \Lambda \big \langle \rho_+ \rho_0^2 \big\rangle  +\epsilon \left(1-\frac{1}{N}\right)\langle \rho_+ \rangle
\end{array} \right. 
\label{stochastic_system}
\end{equation}
where $\Gamma:= -\big[2\epsilon+ (2p-1)\big]$ and $\Lambda := 2(2p-1)$.
The mean-field limit for the evolution of the \emph{averaged} fraction of righttists $\langle \rho_+ \rangle$ in Eq. \eqref{mean_field_limit} is recovered by assuming statistical independence of  the densities $\rho_+$  and $\rho_0$, resulting in
\begin{equation}
    \frac{d}{dt}\langle \rho_+ \rangle = (2p-1)\langle \rho_+  \rangle \langle\rho_0 \rangle-\epsilon \langle \rho_+ \rangle 
\end{equation}
from which the mean-field fixed point in Eq. \eqref{invariance} with $\langle\rho_L \rangle_*>0$   is obtained 
\begin{equation}
    \langle\rho_0 \rangle_* = \frac{\epsilon}{2p-1}
    \label{repeat}
\end{equation}
This assumption of statistical independence also allows us to obtain the stationary value of the second moment  $\big\langle \rho_0^2\big\rangle$ from the second equation in Eq. \eqref{stochastic_system}. Indeed,  $\big\langle \rho_0^2\big\rangle$ satisfies
$$-\big[2\epsilon+ (2p-1)\big]\langle  \rho_0 \rangle_* + 2(2p-1)\langle  \rho_0^2 \rangle_*  +\epsilon \left(1-\frac{1}{N}\right) =0$$
Hence,
\begin{equation}
\begin{aligned}
 {\rm Var}(\rho_0) &= \frac{\epsilon}{2(2p-1)}\left[ \frac{2\epsilon}{2p-1}+ \frac{1}{N}\right] -   \frac{\epsilon^2}{(2p-1)^2}\\ &= \frac{\epsilon}{2(2p-1)}\frac{1}{N}
 \end{aligned}
\end{equation}

\section{Derivation of Eq. \eqref{full_master3}}
\label{sec:Derivation3}
We here show how Eq. \eqref{full_master3} can be derived from a master equation for $\mathbb{P}(\mathbf{x},t)$ that represents the distribution of chemical species reacting according to the set of reactions in Eq. \eqref{baseline_reaction}. For every node $i$, we  introduce its local fields:
\begin{equation}
    h^{(0)}_i : = \sum_{j\in\partial_i} \delta_{x_j,0}\,,  \quad h^{(+)}_i:=\sum_{j\in\partial_i}\delta_{x_j,1} \,,  \quad h^{(-)}_i:=\sum_{j\in\partial_i}\delta_{x_j,-1}
\end{equation}
Thus if $i$ has $\kappa_i$ neighbors, then
$h^{(0)}_i + h^{(+)}_i+   h^{(-)}_i =\kappa_i $.
The  master equation for $\mathbb{P}(\mathbf{x},t)$ reads
\begin{equation}
   \frac{1}{N} \frac{d}{dt}\, \mathbb{P}(\mathbf{x}',t) =  \sum_{\{\mathbf{x}\}} \mathbf{W}\big(\mathbf{x}'|\mathbf{x}\big)\,\mathbb{P}(\mathbf{x},t)  - \mathbb{P}(\mathbf{x}',t)
\label{master_equation}
\end{equation}
where, as we consider that only one agent can change its state at any moment in time, the transition rate $\mathbf{W}\big(\mathbf{x}'|\mathbf{x}\big)$ from  $\mathbf{x}:= \big(x_1, x_2,\cdots, x_i, \cdots, x_N\big)$ to $\mathbf{x}':= \big(x_1, x_2,\cdots, x_i', \cdots, x_N\big)$ is given by
\begin{equation}
\mathbf{W}\big(\mathbf{x}'|\mathbf{x}\big):=\frac{1}{N}\sum_{i=1}^N \left[\prod_{j=1(\neq i)}^N\delta_{x_j, x_j'}\right] \mathcal{F}\big(x_i'|x_i\big)
\label{full_rate_matrix}
\end{equation}
with the individual rate matrix $\mathcal{F}\big(x_i'|\{x_i, \mathbf{x}_{\partial_i}\}\big) \equiv \mathcal{F}\big(x_i'|x_i\big)$
\begin{equation}
\mathcal{F}\big(x_i'|x_i \big)=  \begin{pmatrix}  \displaystyle (0|0) & (0|1) & (0|-1)\vspace*{0.1cm} \\ (1|0)  & (1|1) & 0\vspace*{0.1cm} \\ \displaystyle (-1|0) & 0  &  (-1|-1)\vspace*{0.1cm}\end{pmatrix}
\end{equation}
subject to a normalisation constraint:
\begin{equation}
\sum_{x'_i} \mathcal{F}\big(x_i'|x_i\big)= \mathcal{F}\big(x_i|x_i\big) + \sum_{x'_i(\neq x_i) } \mathcal{F}\big(x_i'|x_i\big) = 1    
\end{equation}
and specified explicitly as 
\begin{equation}
  \begin{aligned}
 \mathcal{F}(-1|1) &=0  \,,\qquad  \mathcal{F}(1|-1) =0 \\\mathcal{F}(0|0) &= 1- \frac{k_1h_i^{(+)} + k_4h_i^{(-)}}{\kappa_i} \\
  \mathcal{F}(1|0)& =  \frac{k_1h_i^{(+)} }{\kappa_i}  \,,\qquad  \mathcal{F}(-1|0) =  \frac{k_4h_i^{(-)} }{\kappa_i}  \\
\mathcal{F}(0|1)& = \frac{k_2h_i^{(0)}}{\kappa_i}  +\epsilon \,,\qquad \mathcal{F}(1|1)= 1 -\epsilon -\frac{k_2h_i^{(0)}}{\kappa_i}  \\ \mathcal{F}(0|-1)&= \frac{k_5h_i^{(0)}}{\kappa_i}  +\epsilon   \,,\quad \mathcal{F}(-1|-1)= 1 -\epsilon -\frac{k_5h_i^{(0)}}{\kappa_i}   
\label{all_transitions2}
\end{aligned}
\end{equation}
where $k_1=k_4=p$ and $k_2=k_5=1-p$. Denoting the vector of all nodes' states apart from $i$ as $\mathbf{x}_{\backslash i}$, according to Eq. \eqref{full_rate_matrix}  we have $\mathbf{x}_{\backslash i} = \mathbf{x}'_{\backslash i}$. Now substituting Eq. \eqref{full_rate_matrix} into Eq. \eqref{master_equation}, we obtain the following explicit form of the master equation:
\onecolumngrid

\begin{equation}
\begin{aligned}
  \frac{d}{dt}\, \mathbb{P}(\mathbf{x}',t)& =  \sum_{i=1}^N \sum_{x_i\in\{-1,0,1\}} \left[\mathcal{F}\big(x_i'|x_i\big)\mathbb{P}(\mathbf{x}'_{\backslash i},x_i, t)\right]   - N\mathbb{P}(\mathbf{x}',t) \\ &=  \sum_{i=1}^N 
\mathcal{F}\big(x_i'|x_i'\big)\mathbb{P}(\mathbf{x}', t)
  +\sum_{i=1}^N \sum_{\substack{x_i\in\{-1,0,1\} \\ x_i\neq x'_i}}\mathcal{F}\big(x_i'|x_i\big)\mathbb{P}(\mathbf{x}'_{\backslash i},x_i, t)  - N\mathbb{P}(\mathbf{x}',t) \\ &=  \sum_{i=1}^N 
\Big[1-\sum_{x_i(\neq x_i') } \mathcal{F}\big(x_i|x_i'\big)\Big]\mathbb{P}(\mathbf{x}', t)
  +\sum_{i=1}^N \sum_{x_i(\neq) x_i'}\mathcal{F}\big(x_i'|x_i\big)\mathbb{P}(\mathbf{x}'_{\backslash i},x_i, t)  - N\mathbb{P}(\mathbf{x}',t) \\ &=  \sum_{i=1}^N 
\sum_{x_i(\neq x_i') } \left[\mathcal{F}\big(x_i'|x_i\big)\mathbb{P}(\mathbf{x}'_{\backslash i},x_i, t) - \mathcal{F}\big(x_i|x'_i\big)\mathbb{P}(\mathbf{x}', t)\right]
\label{full_master5}
\end{aligned}
\end{equation}
Multiplying both sides of this equation by $\big[\delta_{x'_i,1}+\delta_{x'_i,-1}\big]$ and following Eq. \eqref{all_transitions2}, $\mathcal{F}\big(1|x_i\big) \neq 0$ and $\mathcal{F}\big(-1|x_i\big) \neq 0$ for $x_i =0$; and 
$\mathcal{F}\big(x_i|1\big) \neq 0$ and $\mathcal{F}\big(x_i|-1\big)\neq 0$ for $x_i =0$, we arrive at 
\begin{equation}
\begin{aligned}
\frac{d\sigma_i}{dt}  
&= \sum_{\{\mathbf{x}'_{\backslash i}\}}\left\{
\Big[\mathcal{F}\big(1|0\big)+ \mathcal{F}\big(-1|0\big)\Big]\mathbb{P}(\mathbf{x}'_{\backslash i},0, t) - \left[\mathcal{F}\big(0|1\big)\mathbb{P}(\mathbf{x}'_{\backslash i},1, t) + \mathcal{F}\big(0|-1\big)\mathbb{P}(\mathbf{x}'_{\backslash i},-1, t)\right]\right\} \\ &= \sum_{\{\mathbf{x}'_{\backslash i}\}}\left\{
p\,\frac{h_i^{(+)}+h_i^{(-)} }{\kappa_i} \mathbb{P}(\mathbf{x}'_{\backslash i},0, t) - \left[\epsilon + (1-p)\,\frac{h_i^{(0)} }{\kappa_i} \right] \Big[\mathbb{P}(\mathbf{x}'_{\backslash i},1, t) + \mathbb{P}(\mathbf{x}'_{\backslash i},-1, t)\Big] \right\}
  \label{full_master2}
  \end{aligned}
\end{equation}
Equation \eqref{full_master2} is exact but unsolvable. This is because, for example, the first term depends on the correlation between $x_i$ and $x_j$, 
\begin{equation}
\sum_{\{\mathbf{x}'_{\backslash i}\}}\Big[h_i^{(+)}+h_i^{(-)}\Big]\mathbb{P}(\mathbf{x}'_{\backslash i},0, t) =  \sum_{\{\mathbf{x}'\}}\mathbb{P}(\mathbf{x}', t) \sum_{j\in \partial_i} [\delta_{x'_i,0}\delta_{x'_j,1}+\delta_{x'_i,0}\delta_{x'_j,-1}\Big]
\end{equation}

Assuming statistical independence  so that $\mathbb{P}(\mathbf{x}'_{\backslash i},0, t) \simeq \mathbb{P}(\mathbf{x}'_{\backslash i}, t) P_i(x_i=0,t)$
\begin{equation}
\sum_{\{\mathbf{x}'_{\backslash i}\}}\Big[h_i^{(+)}+h_i^{(-)}\Big]\mathbb{P}(\mathbf{x}'_{\backslash i},0, t) \simeq  P_i(x_i=0,t) \sum_{\{\mathbf{x}'_{\backslash i}\}} \sum_{j\in \partial_i} [\delta_{x'_j,1}+\delta_{x'_j,-1}\Big] \mathbb{P}(\mathbf{x}'_{\backslash i}, t) = \rho_{i}^{(0)}(t) \sum_{j\in \partial_i} \sigma_{j}(t)
\end{equation}
Similarly, if assuming $\mathbb{P}(\mathbf{x}'_{\backslash i},1, t) \simeq \mathbb{P}(\mathbf{x}'_{\backslash i}, t) P_i(x_i=1,t)$ and $\mathbb{P}(\mathbf{x}'_{\backslash i},-1, t) \simeq \mathbb{P}(\mathbf{x}'_{\backslash i}, t) P_i(x_i=-1,t)$, leads to 
\begin{equation}
\begin{aligned}
    \sum_{\{\mathbf{x}'_{\backslash i}\}} \Big[\mathbb{P}(\mathbf{x}'_{\backslash i},1, t) + \mathbb{P}(\mathbf{x}'_{\backslash i},-1, t)\Big] &= \sigma_i(t) \\
    \sum_{\{\mathbf{x}'_{\backslash i}\}} h_i^{(0)}\Big[\mathbb{P}(\mathbf{x}'_{\backslash i},1, t) + \mathbb{P}(\mathbf{x}'_{\backslash i},-1, t)\Big] &= \sigma_i(t)\sum_{j\in \partial_i} \rho_{j}^{(0)}
    \end{aligned}
\end{equation}
Using all the above expression allows us to arrive at Eq. \eqref{full_master3}:
\begin{equation}
    \frac{d\sigma_i}{dt} =  - \epsilon \sigma_i(t) +\frac{p \rho_{i}^{(0)}}{\kappa_i}\, \sum_{j\in \partial_i} \sigma_j(t)  - \frac{1-p}{\kappa_i}\sigma_i(t)\sum_{j\in \partial_i} \rho_{j}^{(0)}
\label{full_master5} 
\end{equation}
Note that without assuming statistical independence,  the equation for $\sigma_i(t)$ becomes exactly the same as that for $\langle \rho_++ \rho_-\rangle$ in Eq. \eqref{stochastic_system} in the limit of making all particles indistinguishable  (for arbitrarily finite $N$) as $\delta_{x_i,0}  \rightarrow  \rho_0  $, $\delta_{x_i,1}  \rightarrow  \rho_+  $ and $\delta_{x_i,-1}  \rightarrow  \rho_-  $. As in this case, we would have
\begin{equation}
     \frac{d\sigma_i}{dt} =  - \epsilon \sigma_i(t) +\frac{p}{\kappa_i}\, \sum_{j\in \partial_i}\Big[\big\langle \delta_{x_i,0}  \delta_{x_j,1} \big\rangle + \big\langle \delta_{x_i,0}  \delta_{x_j,-1} \big\rangle \Big]  - \frac{1-p}{\kappa_i}\sum_{j\in \partial_i} \Big[\big\langle \delta_{x_i,1}  \delta_{x_j,0} \big\rangle + \big\langle \delta_{x_i,-1}  \delta_{x_j,0} \big\rangle \Big] 
\end{equation}
\twocolumngrid
We first show in Fig. \ref{fig:fig42} that  solutions to Eq. \eqref{full_master3} strongly depend on the initial condition. Specifically, we distinguish between 3 families of initial conditions. In the first one, all the nodes have the same initial state, whose value is generated at random. In the second one, each of the nodes has its own random initial value. For both cases, we can control the ensemble average over the population $\bar{\sigma}(0)=\sum_i \sigma_i(t=0)/N$ by keeping it at a particular value. In the last family of initial condition, we fixed this latter value $\bar{\sigma}(0)$ for a fraction of nodes, while   varying the fraction of initial centrist nodes. Results of this mean-field approach indicate that when the dynamics starts with all nodes having the same values then the system relaxes quickly to a single steady-state. In other cases,  heterogeneity in the initial condition can give rise to multiple stationary states. It would be interesting to investigate in future work what is the probability to end up in any of these final steady state when starting from a given initial condition. 


\section{The $n$-state Ivoter model}
\label{sec:Derivation4}
Let $\rho_{+}$, $\rho_{2+}$, $\rho_{-}$, $\rho_{2-}$ and $\rho_{0}$ denote the densities of voters whose states are $x_i=1$, $x_i=2$, $x_i=-1$, $x_i=-2$ and $x_i=0$, respectively.    The 5-state model $(p,\epsilon, \gamma)$ is given by four extra ODEs: 
\begin{equation}
        \left \{ \begin{array}{l} \displaystyle 
\dot{\rho}_{+} = (2p-1)\Big[\rho_{+}\rho_{0} - \rho_{+}\rho_{2+} -\gamma\rho^2_{+} \Big] + \epsilon\big[\rho_{2+} -\rho_{+}\big]\vspace{3pt}\\ 
\displaystyle  \dot{\rho}_{2+} =  (2p-1)\rho_{+}\rho_{2+} +\gamma\rho^2_{+} -\epsilon\rho_{2+} \vspace{3pt}\\ \displaystyle 
\dot{\rho}_{-} = (2p-1)\Big[\rho_{-}\rho_{0} - \rho_{-}\rho_{2-}\Big] -\gamma\rho^2_{-} + \epsilon\big[\rho_{2-} -\rho_{-}\big]\vspace{3pt}\\ 
\displaystyle  \dot{\rho}_{2-} =  (2p-1)\rho_{-}\rho_{2-} +\gamma\rho^2_{-} -\epsilon\rho_{2-} 
\end{array} \right. 
\label{mean_field_limit_five_state}
    \end{equation}
Therefore, 
\begin{equation}
\begin{aligned}
    d(\rho_{+} + \rho_{2+})/dt &=  (2p-1)\rho_{+}\rho_{0} -\epsilon\rho_{+}\\
     d(\rho_{-} + \rho_{2-})/dt & =  (2p-1)\rho_{-}\rho_{0} -\epsilon\rho_{-}
     \end{aligned}
\end{equation}
The fixed point of the  dynamics for $\rho_0$ in the 5-state model
\begin{equation}
    \dot{\rho}_0 = -\tilde{y}\big[(2p-1)\rho_0 -\epsilon \big]\,,\qquad \tilde{y}= \rho_{+} +\rho_{-}
    \label{mean_field_5state}
\end{equation}
is  the same as in Eq. \eqref{invariance} for $p>1/2$ and is independent of $\gamma$. For the 5-state model with the two additional states $+2$ and $-2$, the set of reactions includes the following additional reactions with $k_7=k_8  = \gamma$
  \begin{equation}
        \left \{ \begin{array}{l} \displaystyle L_2+ L \overset{k_1}{\longrightarrow} L_2+L_2
\vspace{3pt}\\ 
\displaystyle   L_2+L \overset{k_2}{\longrightarrow} L+L \vspace{3pt}\\ 
\displaystyle   L_2\overset{k_3}{\longrightarrow} L \vspace{3pt}\\ 
\displaystyle   R_2+R \overset{k_4}{\longrightarrow} R_2+R_2 \vspace{3pt}\\ 
\displaystyle   R_2+R \overset{k_5}{\longrightarrow} R+R \vspace{3pt}\\ 
\displaystyle   R_2 \overset{k_6}{\longrightarrow} R \vspace{3pt}\\ 
\displaystyle   R+R \overset{k_7}{\longrightarrow} R_2 +R_2 \vspace{3pt}\\ 
\displaystyle   L+L \overset{k_8}{\longrightarrow} L_2 + L_2\vspace{3pt}\\ 
\end{array} \right. 
    \end{equation}
   The   Hamiltonian in this case  reads
    \begin{equation}
\begin{aligned}
    H = &~ p\Big[\big( a_L^\dag \big)^2 - a_L^\dag a_C^\dag  \Big]a_L a_C +k_2\Big[\big( a_C^\dag \big)^2 - a_L^\dag a_C^\dag  \Big]a_L a_C \\ &+ p\left\{\Big[\big( a_R^\dag \big)^2 - a_R^\dag a_C^\dag  \Big] +k_2\Big[\big( a_C^\dag \big)^2 - a_R^\dag a_C^\dag  \Big]\right\}a_R a_C  \\ &+
    p\left\{\Big[\big( a_{L_2}^\dag \big)^2 - a_L^\dag a_{L_2}^\dag  \Big] +k_2\Big[\big( a_L^\dag \big)^2 - a_L^\dag a_{L_2}^\dag  \Big]\right\}a_L a_{L_2} \\ &+
    p\left\{\big[( a_{R_2}^\dag )^2 - a_R^\dag a_{R_2}^\dag  \big]+k_2\big[( a_R^\dag)^2 - a_R^\dag a_{R_2}^\dag  \big]\right\}a_R a_{R_2} \\ &+
    \gamma \Big[ \big(a_{R_2}^\dag\big)^2  - \big(a_R^\dag\big)^2  \Big]\big(a_R\big)^2 +
    \gamma \Big[ \big(a_{L_2}^\dag\big)^2  - \big(a_L^\dag\big)^2  \Big]\big(a_L\big)^2  \\ &+ \epsilon \Big[ a_C^\dag - a_L^\dag   \Big]a_L + \epsilon \Big[ a_C^\dag - a_R^\dag   \Big]a_R  \\ &+ \epsilon \Big[ a_{L}^\dag - a_{L_2}^\dag   \Big]a_{L_2} + \epsilon \Big[ a_{R}^\dag - a_{R_2}^\dag   \Big]a_{R_2} 
    \end{aligned}
\end{equation}
from which, we can obtain the equation of motion for $\langle n_L \rangle$, $\langle n_{L_2} \rangle$ and $\langle n_L n_C \rangle$:
\begin{equation}
 \begin{aligned}
    \frac{d}{dt} \langle n_L\rangle = & ~ (2p-1)\big(\langle  n_Ln_C \rangle  - \langle  n_Ln_{L_2} \rangle \big) + \epsilon \big(\langle n_{L_2}\rangle  - \langle n_L\rangle \big)\\ &- 2\gamma\big \langle n_L(n_L-1 )\big\rangle
\\ 
 \frac{d}{dt} \langle n_{L_2}\rangle 
 =   & ~ (2p-1)\langle  n_Ln_{L_2} \rangle  -\epsilon \langle n_{L_2}\rangle + 2\gamma \langle n^2_L\rangle -2\gamma \langle n_L\rangle 
\\ 
    \frac{d}{dt}\langle n_L n_C \rangle  = &~  \epsilon\big(\langle n_L n_R \rangle + \langle n_C n_{L_2}\rangle \big) -(1+\epsilon-2\gamma)\langle n_L n_C \rangle\\
    & +(2p-1)\Big[\big\langle n_L n_C^2 \big\rangle -\langle n_L n_R n_C\rangle - \langle n_L n_{L_2} n_C\rangle \Big]\\ &-(2p+2\gamma-1)\big\langle n_L^2 n_C \big\rangle +\epsilon \big\langle n_L(n_L-1) \big\rangle  \label{second_moment_five_state}
\end{aligned}
\end{equation} 
Following similar calculations to what was used after Eq. \eqref{mean_field_limit_five_state} yields
\begin{equation}
    \frac{d}{dt} \langle n_{C}\rangle 
     =-(2p-1)\big\langle  (n_L+n_R)n_{C} \rangle  +\epsilon \langle (n_L+ n_R)\rangle
\end{equation}
Dividing both sides by $N$ as well as assuming statistical independence between $n_C$, $n_L$ and $n_R$ , we arrive at   the same Eq. \eqref{mean_field_5state} for $\langle \rho_{0}\rangle = \langle n_{C}\rangle/N$. This means that the steady-state fraction of centrists $\langle \rho_0\rangle_*$ is  independent of $\gamma$ and equals to that given in Eq.  \eqref{repeat}  if  the set of moment equations is closed at the second order. A similar line of analysis shows that this also holds for the $n$-state model in the mean-field limit.

\begin{figure}
\centering
\includegraphics[width=.6\linewidth]{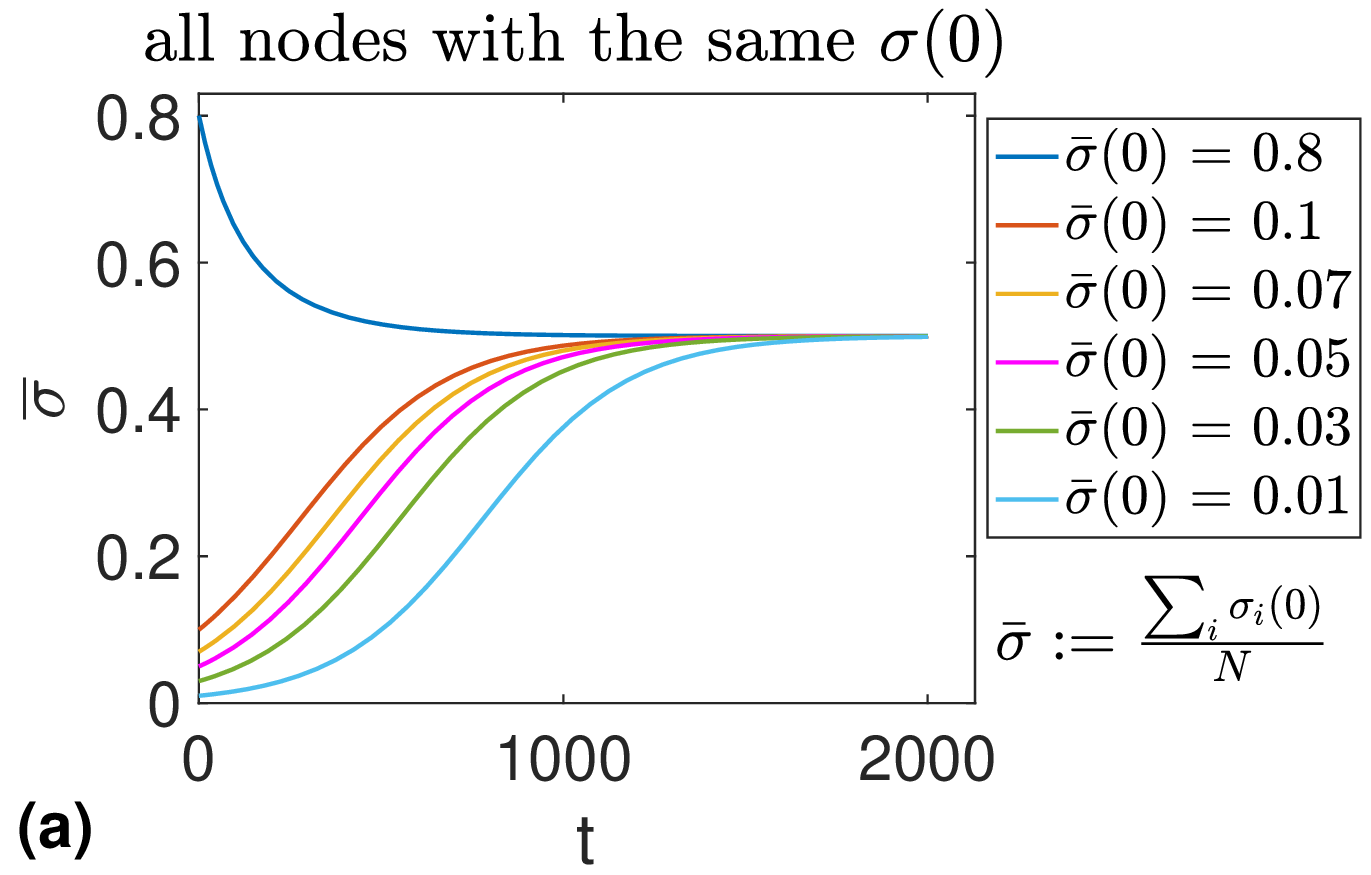}

\includegraphics[width=.6\linewidth]{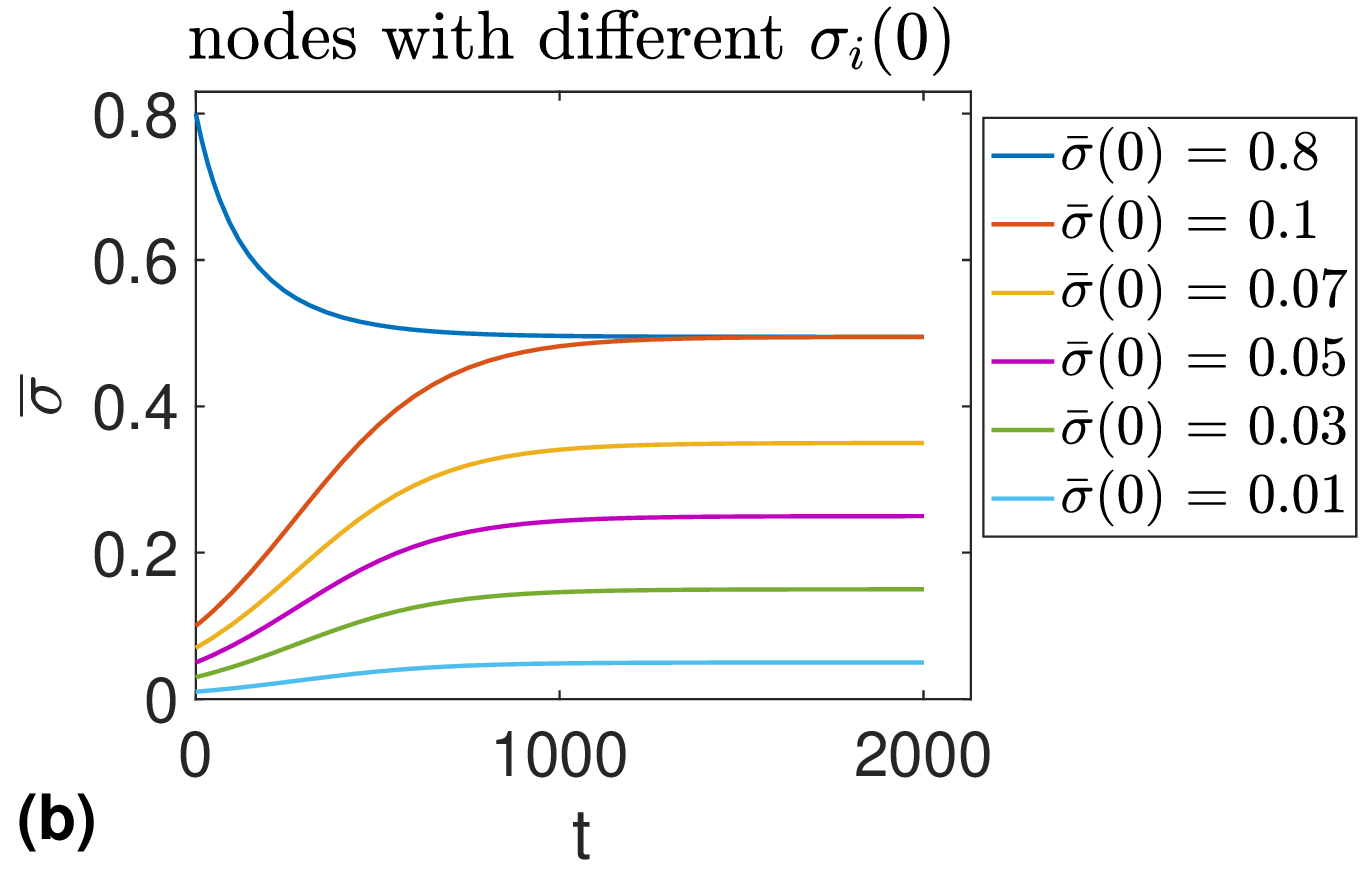}

\includegraphics[width=.6\linewidth]{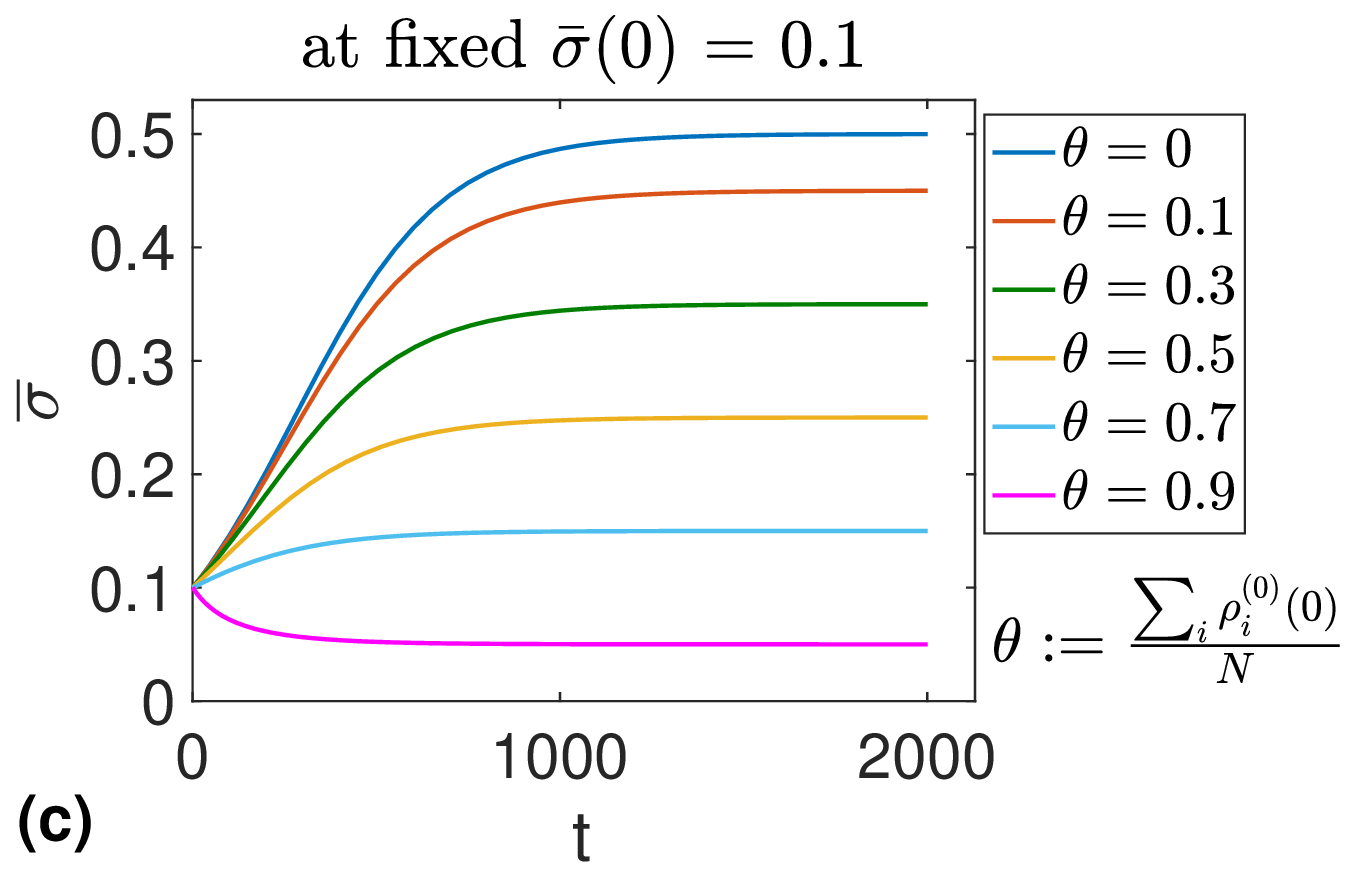}
\caption{The averaged extremeness of the population $\bar{\sigma}(t):= N^{-1}\sum_{i}\sigma_i(t)$ computed from Eq. \eqref{full_master3} for various initial conditions. Here  $N=2000$,  $\kappa =50$,  $\epsilon=0.1$, $p=0.7$.}
    \label{fig:fig42}
\end{figure}

\section{The role of noise} 
\label{sec:noise}
 To test the robustness of our results reported in the main text we introduce a random flip of centrist to either leftist or rightist with probability (per unit time) $\lambda$. So $\lambda$  represents the effect of noise in the system as long as $\lambda \ll \epsilon$. Differently from the noisy voter model \cite{GRANOVSKY1995}, we exclude the spontaneous changes from left to right and vice versa.  Such noise can arise from many different factors that lead to a random  flip of an individual's opinion regardless of the state of its neighbors. The inclusion of $\lambda>0$ also prevents the system from reaching  an absorbing state of all agents being neutral. The individual rate matrix given in Eq. \eqref{all_transitions2} gets modified  as follows:
\begin{equation}
  \begin{aligned}
\mathcal{F}(0|0) &= 1- \frac{k_1h_i^{(+)} + k_4h_i^{(-)}}{\kappa_i} -2\lambda \\   \mathcal{F}(1|0) &=  \frac{k_1h_i^{(+)} }{\kappa_i} +\lambda \,,\quad  \mathcal{F}(-1|0) =  \frac{k_4h_i^{(-)} }{\kappa_i} +\lambda 
\end{aligned}
\end{equation}
Results for fixed  $\lambda=0.05$ on networks of $N=100$ with various values of $\kappa$   are presented in Figure. 6 \textbf{(a)}. Here we confirm that our main result for $\lambda = 0$ (increased polarisation in more connected social networks) is robust wrt the inclusion of $\lambda > 0$.  Next, we test the quality of the MF solution for various $\lambda$ in Figure. 6 \textbf{(b)} and find that it
  agrees better with the simulations as $\lambda$ increases. 
  All curves corresponding to different $\kappa$ merge at high enough $\lambda$, when the effect of noise dominates the I-voter dynamics.

\begin{figure}
\centering
\includegraphics[width=.7\linewidth]{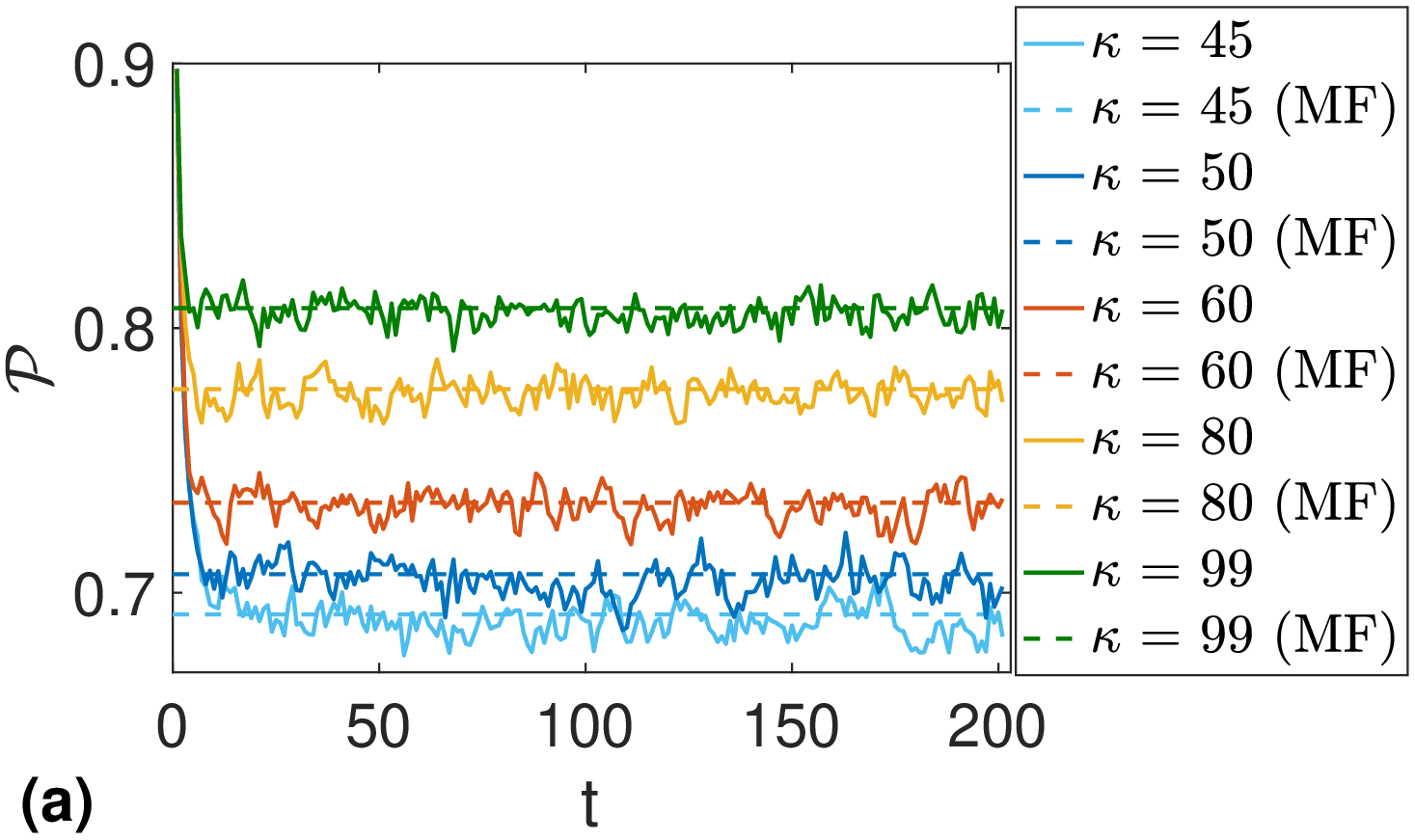} 
\includegraphics[width=.7\linewidth]{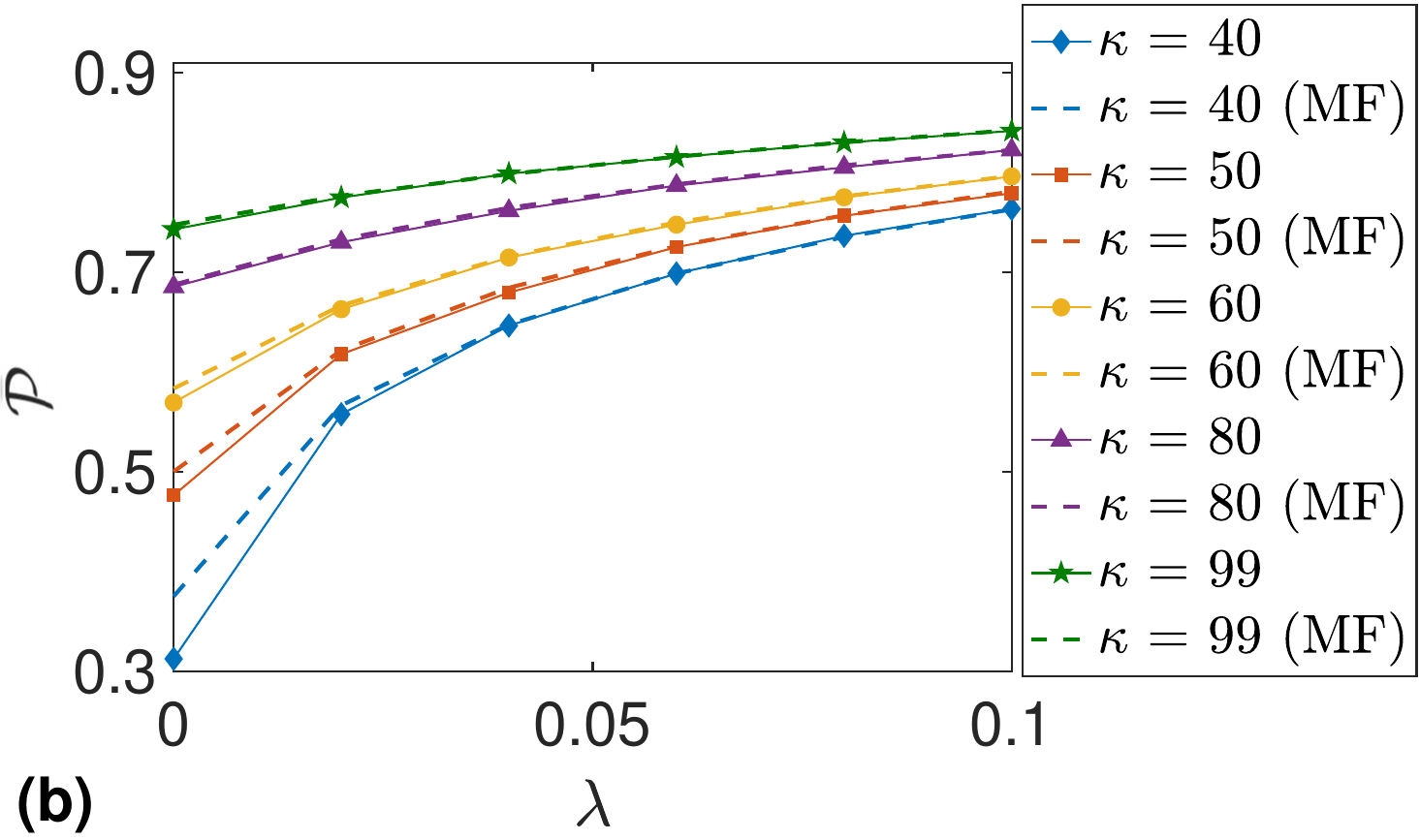} 
\caption{The polarisation measure $\mathcal{P}$ in a social network of varying dergees $\kappa$  with random  flipping of a  centrist to  either leftist or rightist at rate $\lambda = 0.05$ \textbf{(a)} and at different $\lambda$ \textbf{(b)}. Continuous lines are stochastic trajectories generated from the Gillespie algorithm for $N=100$, and then averaging over 100 independent runs. Dashed lines depict the ``MF" prediction according to Eqs. \eqref{full_master4}-\eqref{full_master3}. Here  $\epsilon=0.1$ and  $p=0.7$; the initial fractions of leftists and rightist are equal $0.45$. 
}
\label{fig:fig6}
\end{figure}
\bibliography{Ivoter}
 \end{document}